\documentclass[iop]{emulateapj}
\usepackage{courier}
\usepackage[T1]{fontenc}
\usepackage{multirow}
\usepackage{booktabs}
\usepackage{amsmath}
\usepackage{array}
\usepackage{xcolor}

\newcommand{\code}[1]{\texttt{#1}}
\newcommand{\mesa}{\code{MESA}}
\newcommand\logten{\ensuremath{\log_{10}}}
\newcolumntype{C}[1]{>{\centering\let\newline\\\arraybackslash\hspace{0pt}}m{#1}}

\shorttitle{Type IIb SN Progenitors - I}
\shortauthors{Sravan et al.}

\begin{document}


\title{Progenitors of Type II\MakeLowercase{b} Supernovae: I. Evolutionary Pathways and Rates} 

\author{N. Sravan\altaffilmark{1,2}, P. Marchant\altaffilmark{1}, and V. Kalogera\altaffilmark{1}}


\altaffiltext{1}{Center for Interdisciplinary Exploration and Research in Astrophysics (CIERA) and
Department of Physics and Astronomy, Northwestern University, 2145 Sheridan Road, Evanston, IL 60208, USA}
\altaffiltext{2}{Department of Physics and Astronomy, Purdue University, 525 Northwestern Ave., West Lafayette, IN. 47907, USA}


\begin{abstract}
Type IIb supernovae (SNe) are important candidates to understand mechanisms that drive the stripping of stripped-envelope (SE) supernova (SN) progenitors. 
While binary interactions and their high incidence are generally cited to favor them as Type IIb SN progenitors, this idea has not been tested using models covering a broad parameter space.
In this paper, we use non-rotating single- and binary-star models at solar and low metallicities spanning a wide parameter space in primary mass, mass ratio, orbital period, and mass transfer efficiencies.
We find that our single- and binary-star models contribute to roughly equal, however small, numbers of Type IIb SNe at solar metallicity. 
Binaries only dominate as progenitors at low metallicity. 
We also find that our models can account for less than half the observationally inferred rate for Type IIb SNe at solar metallicity, with computed rates $\lesssim4$\% of core-collapse (CC) SNe. 
On the other hand, our models can account for the rates currently indicated by observations at low metallicity, with computed rates as high as $15$\% of CC SNe. 
However, this requires low mass transfer efficiencies ($\lesssim 0.1$) to prevent most progenitors from entering contact.
We suggest that the stellar wind mass-loss rates at solar metallicity used in our models are too high. Lower mass-loss rates would widen the parameter space for binary Type IIb SNe at solar metallicity by allowing stars that initiate mass transfer earlier in their evolution to reach CC without getting fully stripped.
\end{abstract}







\section{Introduction} \label{s:intro}


CC SNe are explosions marking deaths of stars with zero-age main sequence (ZAMS) masses $\gtrsim 8 M_\sun$ \citep[see e.g.][for a recent review]{2009ARA&A..47...63S}. Depending on the absence or presence of hydrogen lines in the supernova spectrum, SNe are classified into Type I or Type II, respectively. The absence of hydrogen features in a Type I CC SN spectrum is attributed to a progenitor star that lacks its outer hydrogen layers. Type IIb SN progenitors exhibit `mild' stripping of their outer layers, initially exhibiting prominent hydrogen spectral features that weaken and disappear in the weeks following explosion. Type I CC (also known as Type Ibc) and Type IIb SNe are therefore also referred to as SE SNe. 
The mechanisms that drive the stripping and regimes in which they dominate are still open questions. The leading candidates are close binary interactions \citep[e.g.,][]{1992ApJ...391..246P,2010ApJ...725..940Y,2017ApJ...840...10Y}, stellar winds \citep[e.g.,][]{1993ApJ...411..823W,2012A&A...542A..29G,2013A&A...558A.131G, 2017MNRAS.470L.102S}, stellar rotation \citep[e.g.,][]{2012A&A...542A..29G,2013A&A...550L...7G,2013A&A...558A.131G}, and nuclear burning instabilities \citep[e.g.,][]{2011ApJ...741...33A,2015ApJ...811..117S}.

Binary interactions were initially the favored channel to strip stars due to the high observed binarity of Wolf-Rayet (WR) stars \citep{1973IAUS...49..205K}. However, with spectroscopic UV observations indicating strong stellar winds, sufficient to strip stars \citep{1978A&A....63..103C}, they became the preferred channel. The trend is now appearing to be reversing, with binary interactions gaining traction as the preferred formation channel.
This is due to a variety of pieces of evidence. 
First, clumping in stellar winds suggest that currently used mass-loss rates are too high; hot wind mass-loss rates are lower by a factor of $2 - 3$ than those typically used in stellar evolution calculations (\citealt{2014ARA&A..52..487S}; but also see \citealt{2012ApJ...751L..34V}).
Second, recent observations indicate that massive stars are predominantly part of close binary systems \citep{2012Sci...337..444S,2014ApJS..213...34K,2017ApJS..230...15M}. 
Third, observed SE SN rates are too high to be explained solely by single-star evolution \citep{2011MNRAS.412.1522S}.
Fourth, pre-SN masses estimated from light curves indicate lower ZAMS mass for SE SN progenitors than those that produce WR stars via the single-star channel \citep{2016MNRAS.457..328L,2019MNRAS.485.1559P}.
Finally, in many cases circumstellar medium densities for SE SNe inferred from X-ray/radio observations indicate pre-SN mass-loss rates higher than WR winds \citep{2012ApJ...752...17W,2016ApJ...821...57D}, implying additional mass loss processes being active close to CC.

Type IIb SNe are of particular interest in understanding formation channels to SE SNe because of a few reasons.
First, they are the only class within the group that has several (five) identified progenitors\footnote{The progenitor of Type Ib SN iPTF13bvn was identified by \cite{2013ApJ...775L...7C} and confirmed by its disappearance by \cite{2016ApJ...825L..22F}. There is also a candidate for the progenitor of Type Ic SN 2017ein \citep{2018arXiv180301050V, 2018MNRAS.480.2072K}.}. 
Second, there is evidence for the presence of a binary companion to the progenitor in three cases \citep{2014ApJ...790...17F,2014ApJ...793L..22F,2018ApJ...856...83R}.
Finally, Type IIb SNe are quite abundant, accounting for $10-12\%$ of all CCSNe and $30-40$\% of all SE SNe \citep{2011MNRAS.412.1473L, 2011MNRAS.412.1522S,2017PASP..129e4201S}. 

SN 1993J is the prototypical Type IIb SN. 
A progenitor candidate was identified in ground-based pre-explosion images by \citet{1994AJ....107..662A}. They also found that its SED had a blue component which they attributed to either an OB association or a binary companion.
Late-time observations of the region have provided stronger evidence for the presence of a putative companion star \citep{2004Natur.427..129M, 2014ApJ...790...17F}.
SN 1993J is the best-studied Type IIb SN to date, in part due to its proximity, being the subject of several observational and theoretical investigations.
Since 1993, putative progenitors of four more Type IIb SNe have been identified in pre-explosion images: SN 2008ax \citep{2008MNRAS.391L...5C,2015ApJ...811..147F}, SN 2011dh \citep{2011ApJ...739L..37M,2011ApJ...741L..28V,2013ApJ...762...74B}, SN 2013df \citep{2014AJ....147...37V,2015ApJ...807...35M}, and SN 2016gkg \citep{2017MNRAS.465.4650K,2017ApJ...836L..12T,2018Natur.554..497B}. 
There is also evidence for binary companions to the progenitors of SN 1993J \citep{2014ApJ...790...17F}, SN 2001ig \citep{2018ApJ...856...83R}, and SN 2011dh \citep{2014ApJ...793L..22F}.
The Galactic supernova remnant, Cassiopeia A, is known to be the result of a Type IIb SN from light echo spectra \citep{2008Sci...320.1195K,2011ApJ...732....3R}. 
However, there is no companion even at deep limits \citep{2017arXiv171100055K,2018MNRAS.473.1633K}.

With the discovery of more events, there were efforts, both observational and theoretical, to understand the population of Type IIb SNe. On the observational side,
\cite{2010ApJ...711L..40C} studied a sample of Type IIb SNe and suggested that they can be further classified into two sub-types based on their radio SN shock velocities; compact and extended Type IIb SNe progenitors exhibit high and low velocities, respectively.
However, this suggestion was quickly challenged by the discovery of SN 2011dh exhibiting both rapidly expanding radio shells \citep{2012ApJ...752...78S} and an extended yellow supergiant progenitor \citep{2014ApJ...793L..22F}. 
\citet{2014ApJ...792....7F} suggested that some Type Ib/c SNe were spectroscopically more similar to Type IIb SNe. More recently, \citet{2016ApJ...827...90L} found a continuum in the signatures of Type IIb and Ib SN spectra.

Most early theoretical investigations into progenitors of and evolutionary pathways to Type IIb focussed on SN 1993J \citep[e.g.,][]{1993Natur.364..509P, 1994ApJ...429..300W, 2004Natur.427..129M, 2009MNRAS.396.1699S}.
\citet{2010ApJ...725..940Y} and \citet{2011MNRAS.414.2985D} studied progenitors of Type Ib/c SNe arising as a result of mass transfer in close binary systems and found that some of their Type Ib SN progenitors 
exploded with small amounts of residual hydrogen. They suggested that these progenitors may be classified as Type IIb SNe if detected soon after explosion. \citet{2012A&A...538L...8G} and \cite{2013A&A...550L...7G} used solar-metallicity single-star, non-rotating and rotating models and identified their $20-25M_\sun$ rotating models as potential Type IIb SN progenitors. 

\citet{2011A&A...528A.131C} performed the first parameter space search for single and binary progenitors of Type IIb SNe using detailed evolutionary calculations. 
However, they restricted their binary parameter space to initial primary masses $15 M_\sun$, initial secondary masses $10 M_\sun-15 M_\sun$, initial orbital periods $800 - 2100$ days, and solar metallicity. 
While they were able to analyze various evolutionary pathways to Type IIb SNe and their outcomes, they were unable to compute robust relative rates because of their limited parameter space coverage. As we show in this paper, the parameter space for binary Type IIb SNe varies significantly with initial primary mass. Another limitation is that they restrict their analysis to progenitors that explode with $0.1-0.5 M_\sun$ of residual hydrogen envelope. This excludes the group of more compact Type IIb SN progenitors suggested from analyses of Type IIb SN light curves \citep{2014MNRAS.445.1647M,2017ApJ...836L..12T,2018Natur.554..497B} and detailed non-LTE radiation hydrodynamical calculations \citep{2018A&A...612A..61D}. 
Recently, \citet{2017ApJ...840...10Y} undertook a wide parameter space search for binary Type IIb and Ib SN progenitors using detailed evolutionary calculations, varying the initial primary-star mass from $10 - 18 M_\sun$, initial orbital period from $10 - 3000$ days, but keeping the initial mass ratio (= 0.9) and mass transfer efficiency (= 0.2) constant. They analyze Type IIb SN progenitors at two different metallicities, $Z =0.007$ and solar, and show that the parameter space for Type IIb SNe broadens significantly when lowering metallicity and that its effect is roughly analogous to lowering the wind mass-loss rate. However, their parameter space coverage precludes them from computing case C Type IIb SNe (mass transfer after core helium exhaustion), Type IIb SN relative rates, and statistical progenitor properties.

Motivated by these gaps in theoretical analyses, in a two paper study, we investigate the progenitors (their evolutionary pathways, rates, and properties) of Type IIb SNe (henceforth referred to as SNe IIb) using a comprehensive parameter space coverage and statistical analysis. Our study yields a comprehensive database of full evolutionary histories of non-rotating single- and binary-star progenitors of SNe IIb at solar and sub-solar metallicities.  
This paper is dedicated to investigating the parameter space and evolutionary pathways to SNe IIb. 
In addition, our parameter space coverage allows us to compute theoretical SN IIb rates. 
The second paper will be dedicated to investigating the observable properties of SN IIb progenitors presented here.

This paper is organized as follows. In Section \ref{s:models}, we provide a detailed description of our models.
In Section \ref{s:channels}, we discuss key evolutionary channels, and governing physics, towards SNe IIb.
In Section \ref{s:pspace}, we describe the parameter space for SN IIb progenitors at solar and sub-solar metallicities. 
In Section \ref{s:rates}, we provide theoretical rates for SNe IIb as a fraction of CC SNe. 
We discuss our results and conclude in Section \ref{s:conclusions}.
We present numerical tests in the Appendix.

\section{Stellar Models} \label{s:models}

We use Modules for Experiments in Stellar Astrophysics \citep[\mesa\ Release 9575;][]{2011ApJS..192....3P,2013ApJS..208....4P,2015ApJS..220...15P,2018ApJS..234...34P} to compute a large grid of non-rotating single- and binary-star models at solar ($Z_\sun$) and 1/4 solar (which we henceforth refer to as `low metallicity') metallicities. 
We choose the latter to represent nearby low-metallicity environments, i.e. between the Large and Small Magellanic Cloud metallicities.
We adopt the value\footnote{We note that the exact value of $Z_\sun$ is not settled \citep[see e.g.,][]{2009ARA&A..47..481A, 2017ApJ...839...55V}.} of $Z_\sun= 0.02$ to allow comparison of our results with earlier studies.
We assume that helium abundance increases linearly from its primordial value Y = 0.2477 \citep{2007ApJ...666..636P} at Z = 0.0 to Y = 0.28 at Z = 0.02 \citep{2011A&A...530A.115B}
and compute radiative opacities using tables from the OPAL project \citep{1996ApJ...464..943I}. 
In the following, we summarize the properties of our models. 

We use the \texttt{basic.net}, \texttt{co\_burn.net}, and \texttt{approx21.net} nuclear networks in \mesa. The \texttt{basic.net} network includes 8 isotopes $^1$H, $^3$He, $^4$He, $^{12}$C, $^{14}$N, $^{16}$O, $^{20}$Ne, and $^{24}$Mg to model both hydrogen and helium burning, with missing isotopes in the PP and CNO chains being accounted for by assuming equilibrium abundances. The \texttt{co\_burn.net} network adds $^{28}$Si as well as various reactions relevant to carbon and oxygen burning, while the isotope network \texttt{approx21.net} includes an $\alpha$ chain up to $^{56}$Ni and is meant to model late burning stages. We have verified that our results are not sensitive to our choice of nuclear network, with larger networks producing equivalent results.
Reaction rates are taken from \citet{1999NuPhA.656....3A} and \citet{1988ADNDT..40..283C}, with preference given to the latter for rates reported in both.

We model convection using the standard mixing-length theory \citep[MLT;][]{1958ZA.....46..108B,1968pss..book.....C}
, adopting the Ledoux criterion, 
with the mixing length parameter, $\alpha_{\rm MLT}$, set to 1.5 which is the \mesa\ default.
Calibrations done for low and intermediate-mass stars indicate values closer to $\alpha_{\rm MLT}=2$ (cf. \citealt{2016A&A...586A.119S}), but these differences do not have significant impact on our results; testing some of our SN IIb models using $\alpha_{\rm MLT}=2$ leads to differences in progenitor pre-SN effective temperatures of only $\simeq 10\%$. 
During late stages of massive star evolution, regions in convective envelopes can approach the Eddington limit 
with convective velocities nearing sound speed, which is inconsistent with the assumptions of standard MLT. 
Since the treatment of the physics in these regimes is a subject of active research, we currently employ a different treatment of convection in \mesa, known as MLT++ \citep[Section 7.2 in][]{2013ApJS..208....4P}, that artificially reduces the super-adiabacity in these regions, implying unspecified additional energy transport.

We include step overshooting by extending the hydrogen-burning convective core boundary determined by the Ledoux criterion by 0.335 of $H_{\rm p}$ \citep{2011A&A...530A.115B}. 
This value was calibrated using observations of LMC stars with masses $\sim 15 M_\odot$, making it a more appropriate choice compared to smaller values derived from intermediate mass eclipsing binaries \citep{2015A&A...575A.117S}.
We assume negligible overshooting for all other convective regions.
Semi-convection occurs when a region that is unstable according to the Schwarzschild criterion is stabilized by a composition gradient.
It has an important yet ill-constrained effect on the evolution of massive stars \citep{1991A&A...252..669L}. 
\mesa\ uses the formulation of \citet{1983A&A...126..207L} to model semi-convection. We adopt the value of the dimensionless free-parameter $\alpha_{\rm sc}$ to be 1.0 \citep{2006A&A...460..199Y}.
Similarly, thermohaline mixing 
can also cause additional mixing by rendering a region that is stable according to the Ledoux criterion 
unstable due to a negative composition gradient. This phenomenon has an important effect on the evolution of accretors in close binary systems \citep[e.g.,][]{2007A&A...464L..57S}. 
\mesa\ uses the formulation of \citet{1980A&A....91..175K} to model thermohaline mixing. We adopt the value of the dimensionless free-parameter $\alpha_{\rm th}$ to be 1.0.

We use luminosity, effective temperature ($T_{\rm eff}$), surface hydrogen mass fraction ($X_{\rm surf}$), and metallicity dependent stellar winds.
When $T_{\rm eff} \geq 1.1\times 10^4\,$K we adopt the prescription of \citet{2001A&A...369..574V} when $X_{\rm surf}$ $\geq$ 0.4 and that of \citet{2000A&A...360..227N} otherwise.
When $T_{\rm eff} \leq 10^4\,$K we adopt the prescription of \citet{1988A&AS...72..259D} scaled by $(Z/Z_\sun)^{0.85}$, to match the metallicity scaling of \citet{2001A&A...369..574V}, where $Z$ is the initial metallicity of a model. For comparison, the prescription of \citet{2000A&A...360..227N} scales as the square-root of surface metallicity.
When $10^4\,$K $< T_{\rm eff}< 1.1\times 10^4\,$K we interpolate between the results for $T_{\rm eff} \geq 1.1\times 10^4\,$K and $T_{\rm eff} \leq 10^4\,$K. 
This wind mass-loss receipe is similar to the `Dutch' wind mass-loss scheme in \mesa, except for the metallicity dependence adopted for \citet{1988A&AS...72..259D}.

We use the binary module of \mesa\ to model binary stars.
The mass lost from the primary due to Roche-lobe overflow (RLO) is computed following the prescription of \citet{1990A&A...236..385K} and is transferred to the secondary with an efficiency (ratio of mass accreted by the secondary to the mass transferred via RLO by the primary), $\epsilon$, that we assume to be constant during the entire evolution. We assume that the remaining mass is lost from the vicinity of the accretor as its stellar winds. Stellar winds are assumed to carry away the specific angular momentum of the mass losing stars. 
We require that primaries transfer at least 1\% of their initial mass in RLO to qualify as `binaries'. 
This is to exclude effectively non-interacting binary-star models that largely resemble their single star analogs. We show in Section \ref{s:rates} that the exact criterion for selecting `binaries' does not effect our inferences for progenitor channels and derived rates significantly. 
Finally, we assume all initial orbits to be circular.

\renewcommand{\arraystretch}{1.25}
\begin{deluxetable*}{llcccccc}
\tabletypesize{\scriptsize}
\tablecaption{{{Initial properties of solar- and low-metallicity models}} \label{t:pspace}}
\tablewidth{0pt}
\tablehead{
\colhead{} & \colhead{} & \multicolumn{3}{c}{Solar Metallicity} & \multicolumn{3}{c}{Low Metallicity} \\
\cmidrule(lr){3-5}\cmidrule(lr){6-8}
\colhead{Type} & \colhead{Property} & \colhead{min} & \colhead{max} & \colhead{interval} & \colhead{min} & \colhead{max} & \colhead{interval}}
\startdata
Single Stars & $\logten(M_{\rm ZAMS}/M_\sun)$ & 1.2 & 1.5 & 0.0005 & 1.4 & 1.7 & 0.0005\\
\midrule
					& $\logten(M_{\rm ZAMS,1}/M_\sun)$ & 1.00 & 1.40 & 0.02 & 1.00 & 1.40 & 0.02 \\
\multirow{2}{*}{Binary Stars} & $q_{\rm ZAMS}$ & 0.225 & 0.975 & 0.05 & 0.225 & 0.975 & 0.05\\
					& &&& & 1.0 & 2.6 & 0.1 \\
					& $\logten(P_{\rm orb}/$d$)$ & 2.5 & 3.8 & 0.02 & &+ &\\
					& & & & & 2.7 & 3.7 & 0.02 
\enddata
\tablenotetext{}{{\bf (1)} $M_{\rm ZAMS,1}$: Initial primary mass}
\tablenotetext{}{}
\tablenotetext{}{{\bf (2)} $q_{\rm ZAMS} \equiv M_{\rm ZAMS,2}/M_{\rm ZAMS,1}$: Initial mass ratio}
\tablenotetext{}{}
\tablenotetext{}{{\bf (3)} $P_{\rm orb}$: Initial orbital period}
\end{deluxetable*}

We start the evolution of every star at the zero-age main sequence (ZAMS). We terminate the evolution if any one of these conditions are met: (1) the central carbon mass fraction drops below $10^{-6}$ for solar-metallicity models and $10^{-3}$ for low-metallicity models, in which case we assume the star has reached CC\footnote{After modeling our solar-metallicity models we found we could stop the evolution of our models earlier and used it in our low-metallicity models. See the Appendix where we evolve representative solar- and low-metallicity models to advanced nuclear burning stages.}, (2) if the hydrogen-envelope mass drops below $0.01M_\sun$, in which case we assume the star is stripped and will explode as a Type Ibc SN, or (3), in binaries, the system evolves into contact, in which case we assume a merger ensues. We assume the surface properties of the star at this stage match the pre-SN state. While this is plausible, 
recent work suggests that waves can efficiently transport energy outwards during core neon and oxygen burning stages. This could result in outbursts and large fluctuations in surface properties of the star in the years or months leading to CC \citep{2012MNRAS.423L..92Q,2014ApJ...780...96S,2017MNRAS.470.1642F}. 
 
Regarding convergence of our models, \mesa\ uses adaptative timestep and spatial resolution, and our setup is such that our runs typically take $\sim 5000$ timesteps and $\sim 2000$ mesh points. To test convergence, for some of our SN IIb progenitors, we have verified that approximately doubling both the temporal and spatial resolution does not modify our results.
 
In the following subsections and Table \ref{t:pspace} we summarize the parameter space for our single- and binary-star models at solar and low metallicities. 
All our models including \mesa\ input files needed to reproduce them are available at \url{zenodo.org/record/3332831\#.XSjnelApA3g}. 

\subsection{Solar-Metallicity Models} \label{ss:solar}

We compute solar-metallicity single-star models with initial mass, $\logten(M_{\rm ZAMS}/M_\sun)$ = 1.2 -- 1.5 ($M_{\rm ZAMS}/M_\sun \simeq$ 16 -- 31.5) in intervals of 0.0005 dex. 

We compute solar-metallicity binary-star models with initial primary mass, $\logten(M_{\rm ZAMS,1}/M_\sun)$ = 1.0 -- 1.4 ($M_{\rm ZAMS,1}/M_\sun \simeq$ 10 -- 25) in intervals of 0.02 dex, initial mass ratio, $q_{\rm ZAMS} \equiv M_{\rm ZAMS,2}/M_{\rm ZAMS,1}$ = 0.225 -- 0.975 in intervals of 0.05, and initial orbital period, $\logten(P_{\rm orb, ZAMS}/$d$)$ = 2.5 -- 3.8 ($P_{\rm orb, ZAMS}/$d$ \simeq$ 316 -- 6310) in intervals of 0.02 dex. 
We only consider case B or later mass transfer (i.e. mass transfer after core hydrogen exhaustion) in this work; a systematic investigation of case A mass transfer (i.e. mass transfer before core hydrogen exhaustion) towards SNe IIb is beyond the scope of this paper and would be an interesting line of future investigation. 
As a result, the upper limit on the initial primary mass is set to the mass of the most massive single-star model that is stripped by its own stellar winds ($M_{\rm ZAMS} = 25.6M_\sun$), since case B or later mass transfer will only result in additional mass loss as the core is already established in these models.
We compute the models for $\epsilon$ = 1.0 (fully conservative mass transfer), 0.5, 0.1 and 0.01. 

Models with $\epsilon$ = 0.5 and 0.1 are the same as those analyzed in \citet{2018ApJ...852L..17S}, where we show that they successfully reproduce the observed properties of the progenitor of SN 2016gkg derived from pre-explosion photometry and its light curve.

\subsection{Low-Metallicity Models} \label{ss:low}

We compute low-metallicity single-star models with initial mass, $\logten(M_{\rm ZAMS}/M_\sun)$ = 1.4 -- 1.7 ($M_{\rm ZAMS}/M_\sun \simeq$ 25 -- 50) in intervals of 0.0005 dex.
Within this mass range only models with initial masses $>40M_\odot$ lose enough of their hydrogen envelopes to qualify as SN IIb progenitors, but at low metallicities these are expected to directly collapse into black holes rather than explode as SNe \citep{2003ApJ...591..288H}.

We compute low-metallicity binary-star models with the same initial primary masses and initial mass ratios as at solar. 
However, we compute a broader range in initial orbital period than at solar metallicity, $\logten(P_{\rm orb, ZAMS}/$d$)$ = 1.0 -- 3.7 ($P_{\rm orb, ZAMS}/$d$ \simeq$ 10 -- 5012, because of the wider parameter space that leads to SNe IIb at low metallicity.
In addition, we use coarse (0.1 dex) and fine (0.02 dex) intervals for initial orbital periods below and above $\logten(P_{\rm orb, ZAMS}/$d$)$ = 2.7 ($P_{\rm orb, ZAMS} \simeq$ 501 days), respectively, to reduce computational demand for modeling short orbital period binaries (experiencing early case B mass transfer) while adequately resolving long orbital period binaries (experiencing late case B and case C mass transfer). 
As for solar metallicity, our coverage of the range in orbital periods does not capture case A mass transfer. We limit the initial primary mass to $M_{\rm ZAMS,1} \simeq 25M_\sun$, even though our single-star models at low metallicity retain large envelopes at this mass and can potentially lead to SNe IIb after mass transfer, to reduce computational demand. In addition, we show in Section \ref{s:rates} that potential SN IIb progenitors with $M_{\rm ZAMS,1} \gtrsim 25M_\sun$ are relatively very few and do not affect our inference for SN IIb progenitors significantly.
We compute all the aforementioned low-metallicity models for $\epsilon$ = 0.5 and 0.1. 
We do not use $\epsilon =$ 1.0 and 0.01, because, as we show in Section \ref{s:pspace}, the parameter spaces at $\epsilon = 0.01$ and 0.1 are quite similar and the parameter space at $\epsilon=1.0$ is quite small at solar metallicity. In addition, mass transfer is expected to be non-conservative due to rapid spin-up of the secondary to critical rotation during mass transfer \citep[][also see \citet{1991ApJ...370..604P} for a counter argument]{1981A&A...102...17P,2005A&A...435.1013P,2012A&A...537A..29R}. 

\subsection{Model Definitions} \label{s:definitons}

We introduce a couple of definitions to aid our discussion of SN IIb progenitors.

First, we define which pre-CC models are interpreted as SN IIb progenitors. The connection between the pre-SN progenitor structure and the explosion's spectroscopic properties is a subject of ongoing investigation \citep[e.g.,][]{2018A&A...612A..61D}. 
Therefore, here we adopt a conservative criterion using the pre-SN hydrogen envelope mass to define a SN IIb progenitor.
Specifically, models with primaries that reach CC with residual hydrogen envelope mass\footnote{The hydrogen envelope-helium core boundary is defined as the outermost point where the hydrogen mass fraction $\leq 0.01$ and the helium mass fraction $\geq 0.1$.} $0.01M_\sun \leq M_{\rm H~env,preSN(,1)} \leq 1M_\sun$ are defined as SN IIb progenitors. This is our `fiducial SN IIb progenitor definition'.

Our adopted definition was chosen to be a conservative one. 
Inferred hydrogen envelope masses for SNe IIb with detected progenitors are $\lesssim 0.5 M_\odot$ \citep{1994ApJ...429..300W,1996ApJ...456..811H,2012ApJ...757...31B,2014MNRAS.445.1647M,2017ApJ...837L...2A,2018Natur.554..497B}.
These SNe IIb represent those with the most massive, and thus most extended \citep{2017ApJ...840...10Y}, envelopes, as more compact progenitors would be harder to detect. 
In addition, the inferred hydrogen envelope masses from a larger population SNe IIb are also $< 0.5 M_\odot$ \citep{2014MNRAS.439.1807B,2015MNRAS.454...95M,2019arXiv190309262F}.

We also test the effect of changing the `fiducial definition' of SN IIb progenitors in later sections by applying a tighter cut on the residual hydrogen envelope mass ($0.01M_\sun \leq M_{\rm H~env,preSN(,1)} \leq 0.5M_\sun$) and the helium core mass \citep[$2M_\sun \leq M_{\rm He~core,preSN(,1)} \leq 6M_\sun$;][]{1993Natur.364..507N,1994ApJ...429..300W,2014MNRAS.445.1647M,2014A&A...562A..17E,2015ApJ...811..147F,2015A&A...580A.142E}. These constraints are motivated by values derived for SNe IIb with detected progenitors, computed by analyzing their light-curves numerically (using hydrodynamical models), analytically or semi-analytically.

Second, in addition to case C binaries defined earlier, we introduce a further distinction between binaries experiencing case B mass transfer:
\begin{itemize}
\item Case early-B (EB): models initiating mass transfer while the primary is crossing the Hertzsprung Gap (HG)\footnote{The HG is defined as the phase after core hydrogen exhaustion to the point where the star dips to a minimum luminosity before rising on the giant branch.}: the time between exhausting hydrogen in its core and the start of its rise on the giant branch 
\item Case late-B (LB): models initiating mass transfer after the primary begins its rise on the giant branch but before it exhausts helium in its core
\end{itemize} 
Once again, case C binaries are those where the primary initiates mass transfer after core helium exhaustion.

\section{Evolutionary Channels of SN II\MakeLowercase{b} Progenitors} \label{s:channels}

\begin{figure}
\includegraphics[width=\columnwidth]{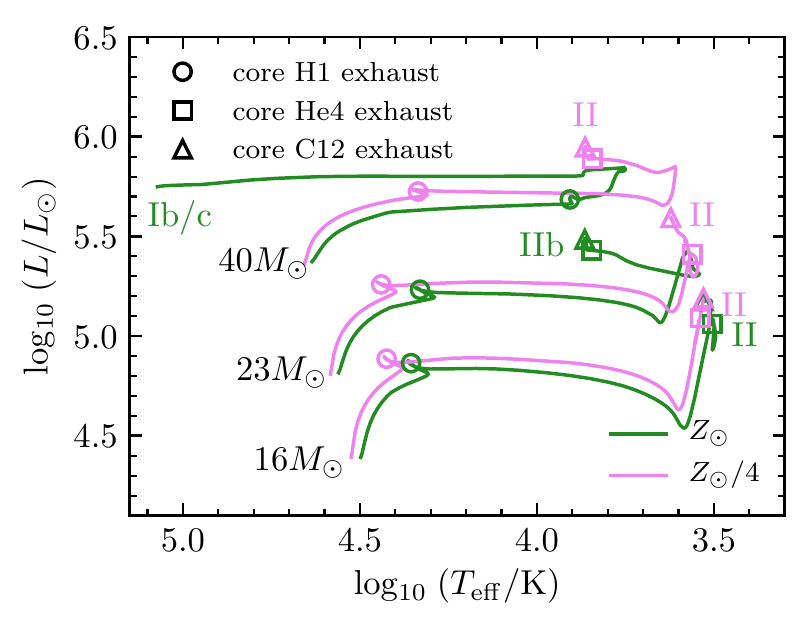}
\caption[]{Hertzsprung-Russell (H-R) diagram showing the evolution of single stars at solar and low metallicities. 
Circles, triangles, and squares denote points when H1, He4, and C12 (see definition in Sections \ref{s:models}) is exhausted in the center of the star, respectively.
Increasing ZAMS mass and metallicity leads to more stripped progenitors due to stronger winds.
At solar metallicity, increasing ZAMS mass leads to a progression to more stripped SN types: from SN II (16 $M_\sun$) to IIb (23 $M_\sun$) to Ibc (40 $M_\sun$).
At low metallicity though, the effect of stripping due to winds is weak: all low-metallicity stars shown here would explode as SNe II.
Note that we stop the evolution of the 40 $M_\sun$ solar-metallicity model when it strips (see Section \ref{s:models}).
\label{f:single}}
\end{figure}

\begin{figure}
\begin{center}
\includegraphics[width=\columnwidth]{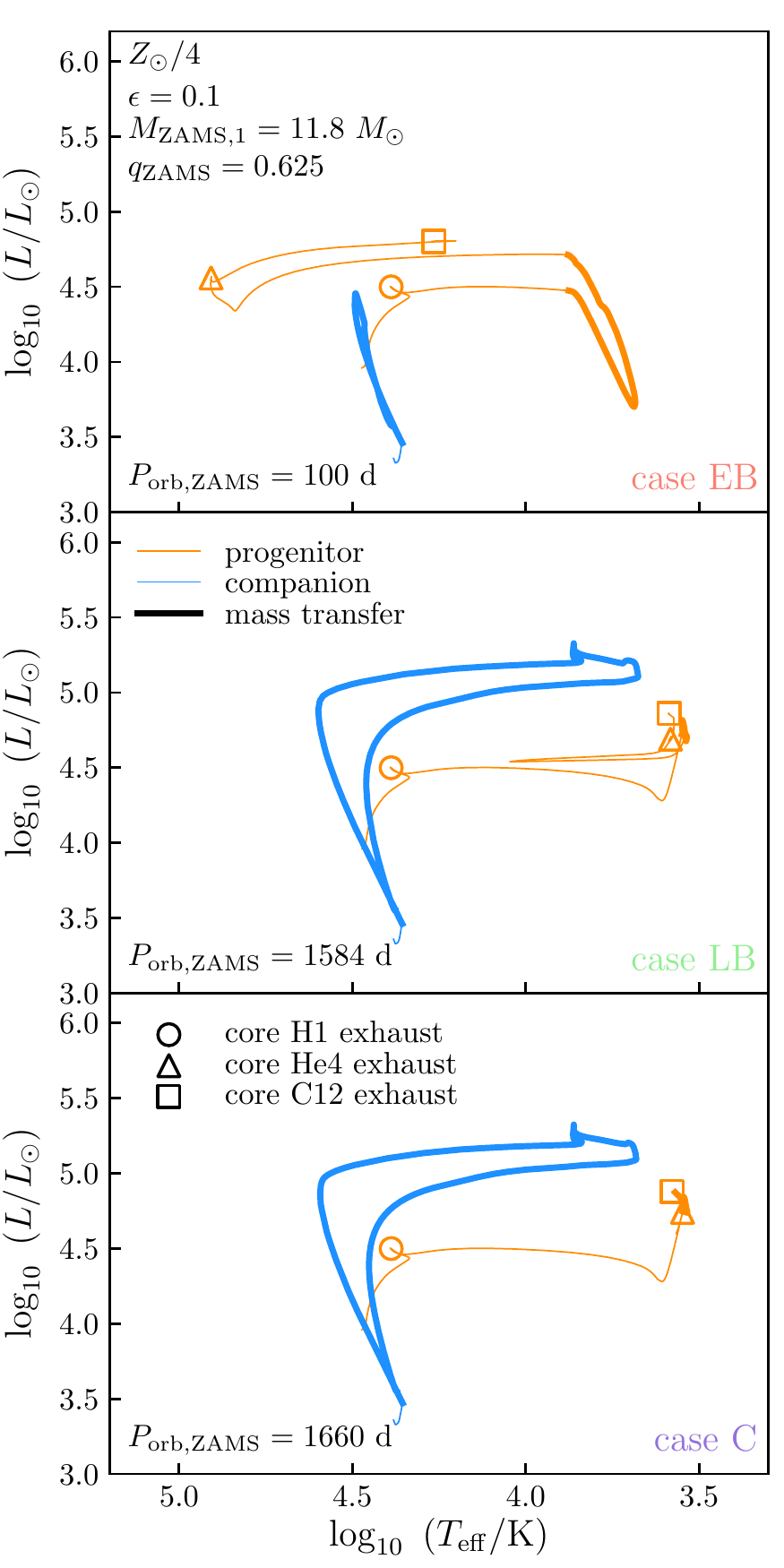}
\end{center}
\caption[]{H-R diagrams showing the three types of evolutionary pathways to binary SNe IIb; via case EB (top), LB (middle), and C (bottom) mass transfer (see Section \ref{s:definitons}). 
The models shown have fixed metallicity (low), mass transfer efficiency (0.1), initial primary mass (11.8 $M_\sun$), and initial mass ratio (0.625) to show the effect of increasing initial orbital period (from top to bottom).
Mass transfer (shown using thicker lines) is defined to be taking place when mass transfer rate due to RLO is $\geq 10^{-6} M_\sun$ yr$^{-1}$.
Circles, triangles, and squares denote points when H1, He4, and C12 (see definition in Sections \ref{s:models}) is exhausted in the center of the corresponding binary component, respectively.
\label{f:binary}}
\end{figure}


\begin{figure*}
\begin{center}
\includegraphics[width=\textwidth]{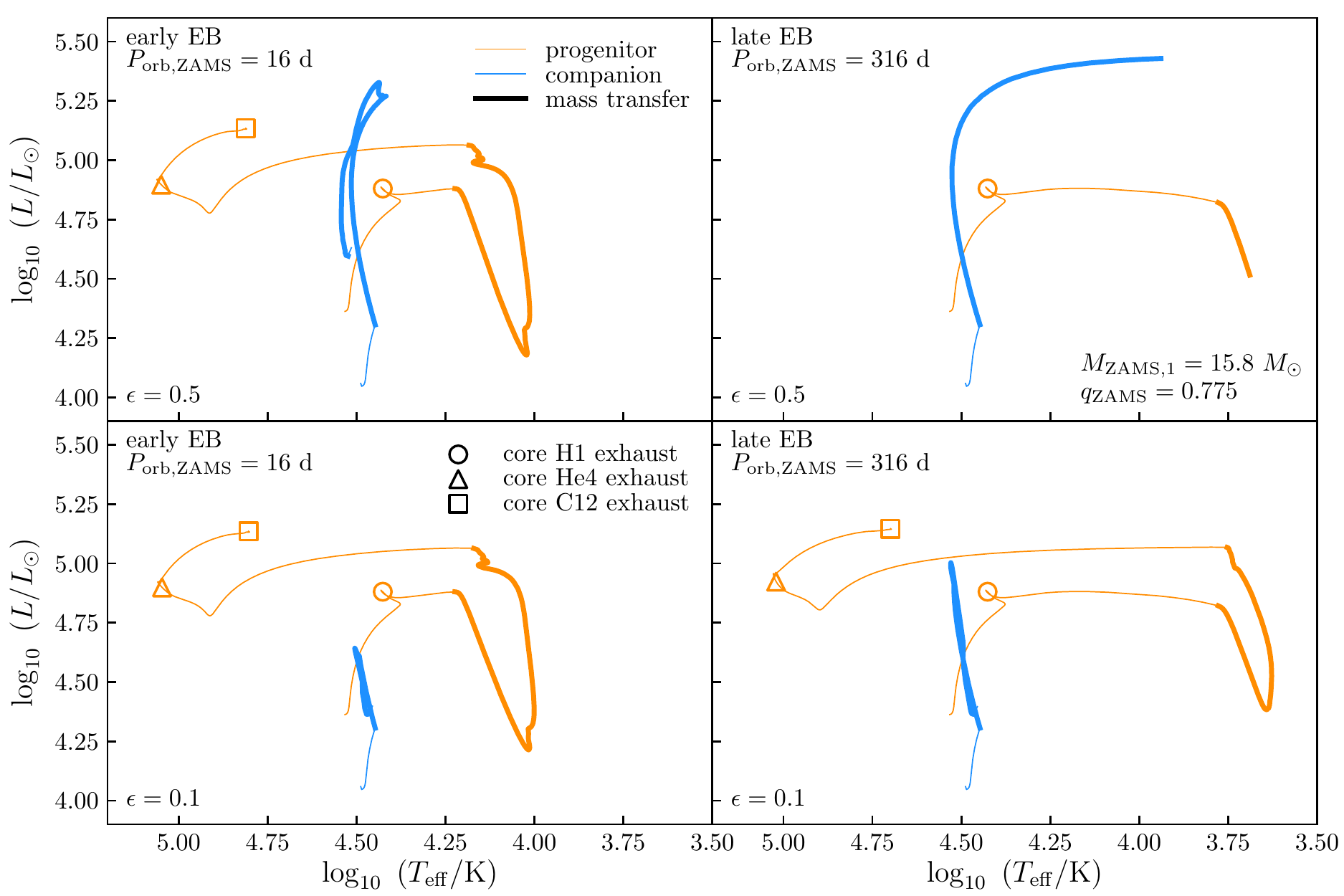}
\end{center}
\caption[]{H-R diagrams showing the difference in evolution of representative low-metallicity binary-star models that initiate mass transfer early (left) or late (right) on the HG due to mass transfer efficiency (0.5, top and 0.1, bottom).
Line colors, line weights, and symbols have the same meaning as in Figure \ref{f:binary}. 
In these models, mass transfer occurs on the thermal timescale of the primary (which is inversely proportional to the primary radius). 
When the primary transfers mass later on the HG (right panels) at higher accretion efficiencies, the secondary approaches a limit, acquiring mass significantly faster than its thermal timescale (model in the top-right panel). This results in the radius of the secondary increasing dramatically and the binary entering a contact phase.
\label{f:gapHR}}
\end{figure*}

In this section, we discuss single and binary evolutionary channels towards SNe IIb at solar and low metallicity,
with a special emphasis on the effect of mass ratio and mass transfer efficiency, especially at low metallicities.
We briefly summarize results from earlier work for context and completeness. Also, we only focus on discussing previous work that used self-consistent stellar evolution models for consistency.

\pagebreak

\subsection{Single SN IIb Progenitors}
 
Figure \ref{f:single} shows H-R tracks for single stars at solar and low metallicity. 
It demonstrates two well-understood phenomena related to producing single stripped stars. First, increasing ZAMS mass drives stronger winds and results in more stripped progenitors. Second, lowering metallicity drives weaker winds and results in less stripped progenitors.

Our single-star solar-metallicity models transition from SN II to SN IIb at ZAMS mass of 22$M_\sun$, then to SN Ibc at ZAMS mass of 26$M_\sun$ (see definitions in Section \ref{s:models} and \ref{s:definitons}). 
In Figure \ref{f:single}, the 23 $M_\sun$ model is a SN IIb progenitor. 
At low metallicity, none of our models lose enough of their envelope to explode as SE SNe.
We emphasize that the mass range leading to a given SN type at any metallicity is strongly dependent on the wind mass-loss prescription used and these are known to be quite uncertain (see also discussion in \ref{ss:physics}).

\subsection{Binary SN IIb Progenitors}

Figure \ref{f:binary} shows typical evolutionary pathways via case EB, LB, and C mass transfer to SNe IIb at low metallicity.
These channels also exist at solar metallicity, however case EB mass transfer channels are strongly suppressed as in several cases stronger winds remove the remaining hydrogen envelope after detachment \citep{2017ApJ...840...10Y}.
Case C mass transfer is only possible for a narrow range in initial primary mass (see also Section \ref{s:pspace}).
This is because only lower primary masses can expand significantly after core helium exhaustion. 
Despite their rarity, case C SNe IIb warrant discussion because they have the attractive property of having recently undergone or still undergoing mass transfer at CC. 
Recent studies indicate that many SNe IIb experience strong mass loss close to CC \citep[e.g., ][]{2016ApJ...818..111K}.
This channel was also initially suggested as the likely progenitor of the prototypical SN IIb: 1993J \citep{1993Natur.364..509P,2004Natur.427..129M}.

Many of these channels were first discussed for solar-metallicity models by \citet{2011A&A...528A.131C}. 
However, unlike this work, \citet{2011A&A...528A.131C} also found channels to SNe IIb via case A and EB mass transfer\footnote{They also did not find channels to SNe IIb via case C mass transfer but this was due to their adopted upper limit on the residual hydrogen envelope mass ($0.5 M_\odot$) for defining SN IIb progenitors.}.
The difference in outcomes is largely due to the adopted wind mass-loss prescription. \citet{2011A&A...528A.131C} adopted the prescription of \citet{1988A&AS...72..259D} during the entire stellar evolution, which is about 2 orders of magnitude lower than our WR mass-loss prescription of \citet{2000A&A...360..227N} and drives most of the stripping in our case.
More recently, \citet{2017ApJ...840...10Y} extended the analysis for SN IIb progenitors to low metallicity. They found a significant difference in the availability of case EB mass transfer channels towards SNe IIb between solar and low metallicities. 
Specifically, though after the mass transfer phase primary stars at both metallicities detach with similar amounts of hydrogen envelope, 
strong winds at high metallicity strip many of them before CC. 
This results in a sharp narrowing of the parameter space for binary SNe IIb at solar metallicity 
as case EB binaries span $\sim$2 orders of magnitude in initial orbital period.
We show the full parameter space of binary SNe IIb as a function of metallicity (including mass transfer efficiency and mass ratio) in Section \ref{s:pspace} and quantify the effect of the above mentioned phenomenon on theoretical SN IIb rates in Section \ref{s:rates}.

Our parameter space coverage allows us to investigate the effect of mass ratio and mass transfer efficiency on binary SN IIb channels at both solar and low metallicities.
We find that the viability of all three channels mentioned above (case EB, LB, and C) increases with increasing initial mass ratio and decreasing mass transfer efficiency.
We find that mass transfer efficiency plays an important role for the viability of case EB SNe IIb.
Left and right panels in Figure \ref{f:gapHR} show the difference in evolution when mass transfer begins early and late on the HG for different $\epsilon$ values.
The primary in these models initiates mass transfer on its thermal timescale, 
which is shorter later on the HG \citep{2001A&A...369..939W}. In late EB mass transfer scenarios (right panels) if $\epsilon$ is high enough that accretion occurs significantly faster than the thermal timescale of the secondary then the binary will enter contact.
Given the wide parameter space spanned by case EB binaries, this phenomenon has an important secondary effect on the parameter space and rates for SNe IIb.

\begin{figure*}
\begin{center}
\includegraphics[width=\textwidth]{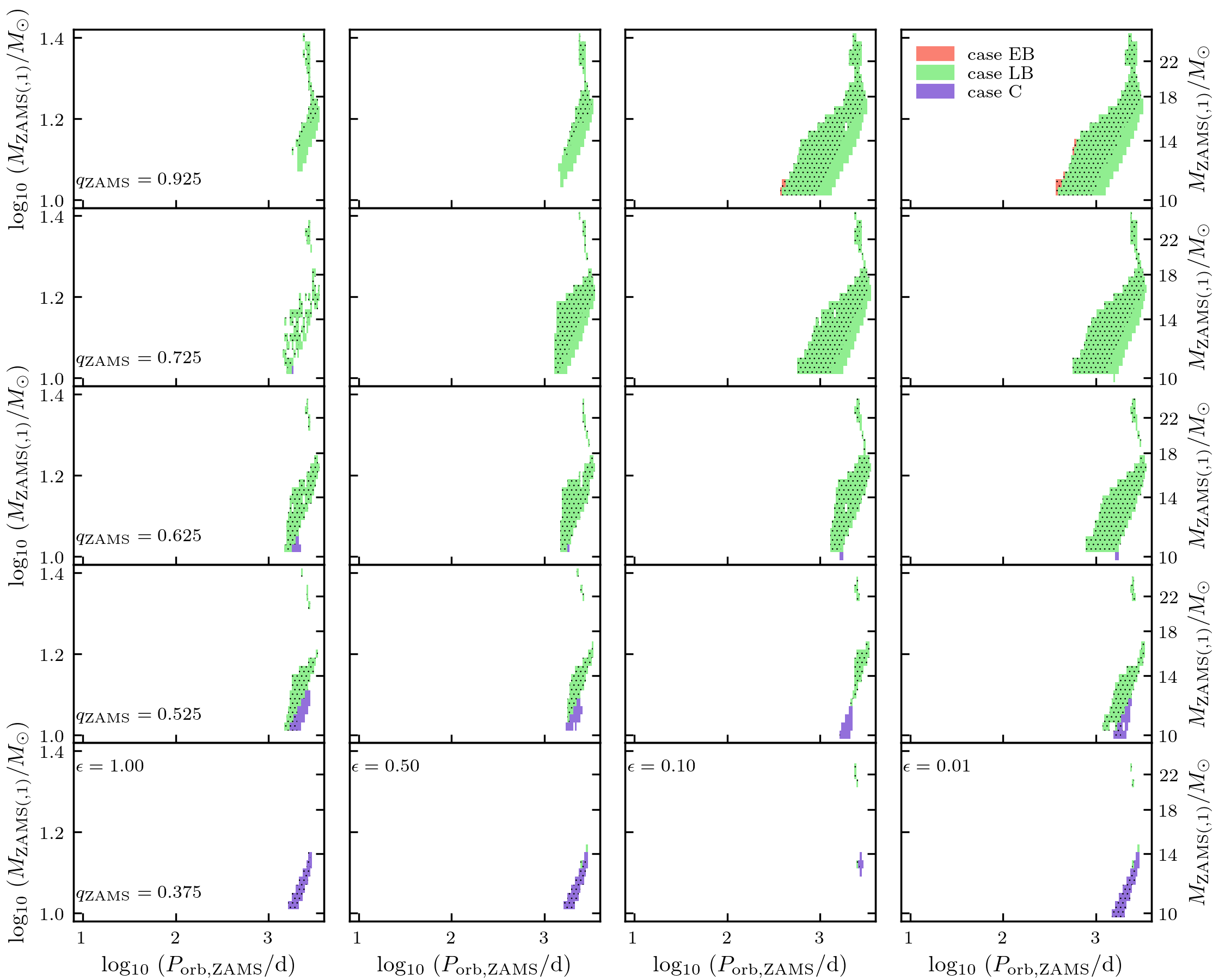}
\end{center}
\caption{Parameter space for binary SN IIb progenitors at solar metallicity. 
Horizontal panels show models with representative $q_{\rm ZAMS} \equiv M_{\rm ZAMS,2}/M_{\rm ZAMS,1}$ (noted to the bottom-left in the left-most panels) demonstrating how the parameter space changes with $q_{\rm ZAMS}$.
Vertical panels from left to right show models with mass transfer efficiency, $\epsilon$ = 1.0 (fully conservative mass transfer), 0.5, 0.1 and 0.01. 
Each model SN IIb is assumed to represent all binaries in its parameter space interval and colored according its mass transfer type (noted in the topmost-right panel).
The dotted regions show the parameter sub-space using the criterion $0.01M_\sun \leq M_{\rm H~env,preSN(,1)} \leq 0.5M_\sun$ to define SN IIb progenitors.
The binary SN IIb parameter space decreases with $q_{\rm ZAMS}$ and decreases more with decreasing $\epsilon$.
\label{f:pspace_solar}}
\end{figure*}

\begin{figure}
\begin{center}
\includegraphics[width=\columnwidth]{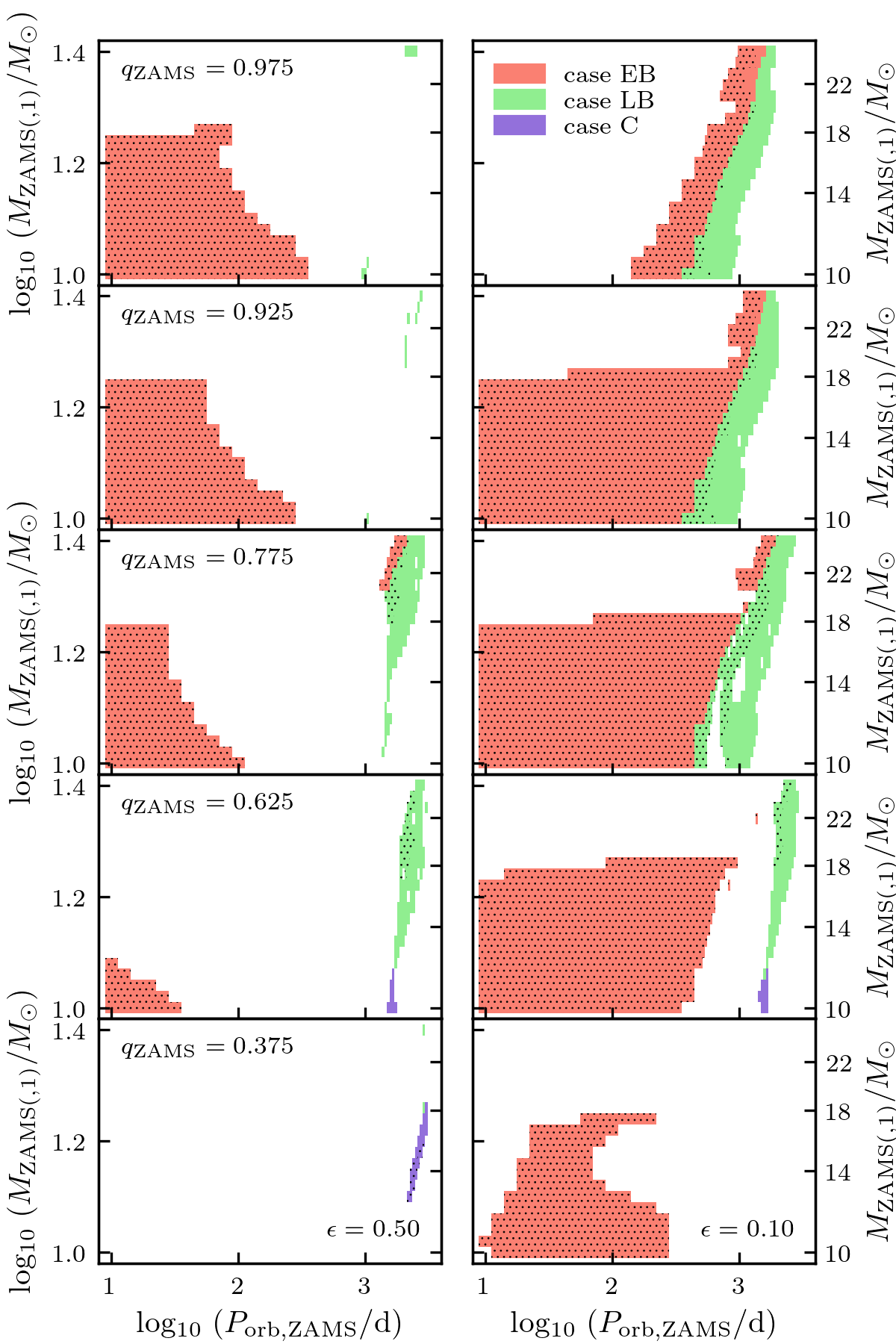}
\end{center}
\caption{Parameter space for binary SN IIb progenitors at low metallicity. Horizontal panels show models with representative $q_{\rm ZAMS}$ (noted to the top-left in the left-most panels) demonstrating how the parameter space changes with $q_{\rm ZAMS}$.
Left and right vertical panels show models with mass transfer efficiency, $\epsilon$ = 0.5 and 0.1, respectively.
The colored and dotted regions have the same meaning as in Figure \ref{f:pspace_solar}.
Case EB binaries dominate the SN IIb population and are favored at low metallicity because weak WR winds fail to strip the hydrogen envelope left over after the mass transfer phase, unlike at solar.
The gap in the parameter space between $\logten(P_{\rm orb, ZAMS}/$d$)\simeq 1.5 (2) -3$ when $\epsilon$ = 0.5 is due to the occurrence of contact in the binaries. 
\label{f:pspace_low}}
\end{figure}

\section{Results on SN II\MakeLowercase{b} Progenitor Populations} \label{s:pspace}





In this section, we discuss the parameter space for single and binary SN IIb progenitors at solar and low metallicities.
For binary systems we focus our discussion on the effect of mass ratio and mass transfer efficiency, especially at low metallicity.

\subsection{Single SN IIb Progenitors} \label{ss:pspsacesingle}

At solar metallicity, single stars with initial mass $\logten(M_{\rm ZAMS}/M_\sun)$ = 1.3445 -- 1.409 ($M_{\rm ZAMS}/M_\sun \simeq$ 22 -- 26) are SNe IIb progenitors according to our fiducial definition (primaries that reach CC with residual hydrogen envelope mass $0.01M_\sun \leq M_{\rm H~env,preSN(,1)} \leq 1M_\sun$).
At low metallicity, our most massive model ($M_{\rm ZAMS} \simeq 50M_\sun$) barely gets stripped enough to be labeled a SN IIb progenitor. 
However, we expect that stars as massive as these (helium core masses $\gtrsim 26 M_\sun$) will not experience successful explosions but will instead collapse directly into black holes \citep{1999ApJ...522..413F}. We adopt this assumption for the remainder of this work.

We also explore the effect of adopting different criteria for defining SN IIb progenitors as the mapping between progenitor structure and ensuing SN spectra is not clear. 
Adopting the cut $0.01M_\sun \leq M_{\rm H~env,preSN(,1)} \leq 0.5M_\sun$ for defining SN IIb progenitors, only models with $M_{\rm ZAMS} \simeq 23 - 25.5~M_\sun$ at solar metallicity qualify as SN IIb progenitors.
Applying the $2M_\sun \leq M_{\rm He~core,preSN(,1)} \leq 6M_\sun$ cut qualifies no single stars, either at solar or low metallicity, as SN IIb progenitors: $M_{\rm He~core,preSN(,1)} = 9.5 M_\sun$ 
for the least massive SN IIb progenitor (using our fiducial definition) at solar metallicity.
This fact has been noted several times in the literature to motivate the need for additional mass-loss mechanisms (including binary interactions) 
to create SN IIb progenitors at lower ZAMS masses \citep[e.g. ][]{2016MNRAS.457..328L,2019MNRAS.485.1559P}.

\citet{2011A&A...528A.131C} found single SN IIb progenitors with ZAMS masses $\sim 33M_\odot$. This is again due to the difference in adopted wind mass-loss prescription, where their formulation is 2 orders of magnitude lower than ours.
Lowering wind mass-loss rates increases the minimum ZAMS mass that produces SNe IIb. 
 
\subsection{Binary SN IIb Progenitors} \label{ss:pspsacebinary}

Figures \ref{f:pspace_solar} and \ref{f:pspace_low} show the parameter space for binary SN IIb progenitors using our fiducial definition (primaries that reach CC with residual hydrogen envelope mass $0.01M_\sun \leq M_{\rm H~env,preSN(,1)} \leq 1M_\sun$) at solar and low metallicity, respectively.

Case LB mass transfer towards SNe IIb are the dominant production scenarios at solar metallicity.
However, case C mass transfer channels, though observationally interesting, have not previously been discussed in the context of a parameter space analysis.
As expected, case C SNe IIb are limited to lower initial primary masses and arise from a small range in initial orbital periods that shrinks with increasing initial primary mass.
Case C SNe IIb are also favored for lower initial mass ratios ($q_{\rm ZAMS} \lesssim 0.65$).

The parameter space for binary SNe IIb increases dramatically with decreasing metallicity. 
This is due to channels to SNe IIb via case EB mass transfer that are only viable at low metallicity (see Section \ref{s:channels}). 
Due to the large change in primary radius that occurs when crossing the HG, the range of initial orbital periods that permit mass transfer during this phase is very large.
The difference in parameter space due to metallicity has a primary dominant effect on theoretically computed SN IIb rates (see Section \ref{s:rates}).



Once again, we explore the effect of adopting different criteria for defining SN IIb progenitors. 
For example, the SN IIb parameter space with $q_{\rm ZAMS} \lesssim 0.4$ is almost entirely progenitors with $M_{\rm H~env,preSN(,1)} \leq 0.5M_\sun$ at both metallicities.
Case EB SNe IIb also dominantly have $M_{\rm H~env,preSN(,1)} \leq 0.5M_\sun$.
Applying the cut $2M_\sun \leq M_{\rm He~core,preSN(,1)} \leq 6M_\sun$, effectively excludes the parameter space above $M_{\rm ZAMS,1} \simeq 16~M_\sun$ at both metallicities. This is because the core is already established before the onset of interaction in the binary models considered in this work (see Section \ref{ss:solar}).

Finally, while it is difficult to directly compare our results to those of \citet{2011A&A...528A.131C} and \citet{2017ApJ...840...10Y} due to differences in modeled binary parameters, definitions for SNe IIb progenitors, and adopted physics in stellar modeling, there is broad agreement with respect to all of the above results.

\subsubsection{Effect of Mass Transfer Efficiency}

Mass transfer efficiency plays an important role in shaping the parameter space towards SNe IIb, especially at low metallicity.
The gap in the parameter space between $\logten(P_{\rm orb, ZAMS}/$d$)\simeq 1.5 (2) -3$ at low metallicity when $\epsilon$ = 0.5 in Figure \ref{f:pspace_low} is due to the occurrence of contact in these binaries (see Section \ref{s:channels} and Figure \ref{f:gapHR} for details).
In other words, low-metallicity EB SN IIb are favored at lower $\epsilon$.
Also overall, the parameter space for SNe IIb decreases with increasing $\epsilon$ at both metallicities. This is because of increasing likelihood of contact in the binaries when the secondary is unable to thermally relax with its acquired mass at higher $\epsilon$ \citep{1995A&A...297..483B} and the decreasing stabilizing effect of orbit widening (from loss of angular momentum) due to inefficient mass transfer.

Interestingly, for the subset of high initial mass ratio ($q_{\rm ZAMS}=0.975$) EB SNe IIb at low metallicity, higher $\epsilon$ has the opposite effect, i.e. it helps avoid contact. 
Specifically, in models with $\epsilon=0.1$, the secondary accretes relatively small amounts of mass when it is close to terminal main-sequence, with negligible effect on its evolution, expands as it leaves the main-sequence and enters contact while on the HG.
However, in models with $\epsilon=0.5$, the secondary accretes enough mass to rejuvenate as a main-sequence star, 
allowing the primary to evolve to CC while the secondary is still on the main-sequence.
This phenomenon results in the absence of several case EB SNe IIb in the topmost-right panel of Figure \ref{f:pspace_low}, as compared to the topmost-left panel of the same figure.

Notably, whether binaries with $\epsilon =$ 1.0 and high mass ratios ($q_{\rm ZAMS} \gtrsim 0.7$) at high metallicity are able to avoid contact depends sensitively on their evolutionary history, which dictates whether convective mixing in the secondary is able to carry the accreted material to the core, allowing them to rejuvenate and contract. 
Specifically, rejuvenation occurs over a series of discrete events where the convective core grows in mass and merges with an outer convection zone. The discrete nature of this process leads to non-monotonic behavior in terms of the initial parameters, resulting in the patchy distribution that can be seen in the panel for $q_{\rm ZAMS} =0.725$ and $\epsilon =$ 1.0 in Figure \ref{f:pspace_solar}.

We note that, the SN IIb parameter space for $\epsilon =$ 0.1 and 0.01 at solar metallicity is quite similar.

\subsubsection{Effect of Mass Ratio}

The parameter space for SNe IIb generally decreases with $q_{\rm ZAMS}$ due to increasing liklihood of contact in the binaries. 
At the onset of mass transfer, if $q=M_{accretor}/M_{donor}\lesssim1$ the orbit to shrinks in response to further mass transfer (with lower mass ratios resulting in faster shrinkage) causing the system to undergo unstable mass transfer. 
However, if $q\gtrsim1$, the orbit expands, stabilizing further mass transfer, allowing the primary to evolve to CC.
The exact boundary that separates the two scenarios depends on the mass transfer efficiency, e.g. for conservative mass transfer it is $q=1$ \citep{2006csxs.book..623T}.
Similarly, the parameter space decreases more with $q_{\rm ZAMS}$ at lower $\epsilon$, as more material leaving the system carrying its orbital momentum causes the orbit to shrink faster, promoting conditions for the development of contact or unstable mass transfer.

However, in binaries with $q_{\rm ZAMS}\lesssim 0.5$, higher values of $\epsilon$ can cause the mass ratio to flip thereby facilitating stable mass transfer. 
This results in the break in the parameter space trend at $\epsilon = 0.1$ in the lowermost panel series of Figure \ref{f:pspace_solar}.
At $\epsilon = 0.1$, several accretors that managed to avoid contact at $\epsilon = 0.01$, overflow their roche lobes. However, at even higher values of $\epsilon$, the initial mass transfer phase causes the mass ratio to rapidly flip, leading to the expansion of the orbit on further mass transfer, allowing them once again to avoid contact.

\section{SN\MakeLowercase{e} II\MakeLowercase{b} Relative Rates} \label{s:rates}

We use models discussed in previous sections to compute SN IIb relative rates (fractions) at solar and low metallicities. 
In the following, we outline our assumptions for the computation of SN rates.

We assume the distribution of the fraction of binary systems, $f_{\rm bin}$, is flat with respect to ZAMS mass, initial mass ratio, and initial orbital period.
The distribution of initial mass, $M_{\rm ZAMS}$, of all stars is assumed to be the Salpeter IMF \citep{1955ApJ...121..161S}
\begin{equation} \label{e:IMF}
f(M_{\rm ZAMS}) = (M_{\rm ZAMS})^{-\alpha} .
\end{equation}
We assume that the minimum ZAMS mass to undergo CC is $8 M_\sun$\footnote{ \citet{2007A&A...476..893S} and \citet{2015MNRAS.446.2599D} suggest a higher value of $\sim10-12 M_\sun$, depending on the choice of convection and mass loss.} \citep{2002RvMP...74.1015W,2009ARA&A..47...63S}. 
Following \citet{2003ApJ...591..288H}, at solar metallicity we do not consider an upper mass limit for stars to explode as SNe, assuming that even stars that form black holes explode as fallback SNe. Also based on the results of \citet{2003ApJ...591..288H}, for our low-metallicity models we consider stars with initial masses in excess of $40M_\odot$ to directly collapse to black holes rather than explode as SNe.
We adopt a power-law distribution for the initial mass ratio, $q_{\rm ZAMS}$, 
\begin{equation} \label{e:f(q)}
f(q_{\rm ZAMS}) = (q_{\rm ZAMS})^\beta .
\end{equation}
We assume that it is valid between $0.2\leq q_{\rm ZAMS}\leq1.0$ \citep{2014ApJS..213...34K}.
Finally, we assume a power-law distribution for the initial orbital period, $\logten P_{\rm orb, ZAMS}$,
\begin{equation} \label{e:f(P)}
f(\logten P_{\rm orb, ZAMS})=(\logten P_{\rm orb, ZAMS})^{\gamma} .
\end{equation}
We assume that it is valid $0.15 \leq \logten (P_{\rm orb,ZAMS}/$d$) \leq 3.80$ (the upper limit on the initial orbital period modeled in this work). Note that while massive binary properties are observed to be relatively insensitive to metallicity only between solar and Large Magellanic Cloud metallicities \citep[and initial orbital periods from 1--1000 days; ][]{2017A&A...598A..84A}, we assume that they are preserved down to a fourth-solar.

We use several values of $f_{\rm bin}$, $\alpha$, $\beta$, and $\gamma$.
For $f_{\rm bin}$ we use 0.25, 0.5, 0.65, and 0.8. 
The first two are to allow comparison of our results against those of \citet{2011A&A...528A.131C}, while the last two are the upper and lower limits on $f_{\rm bin}$ found by \citet{2014ApJS..213...34K} (\cite{2012Sci...337..444S} find $f_{\rm bin} \sim 0.7$ for Galactic O-stars). 
For $\alpha$ we use 1.6, 2.3, 3.0 to capture the 1-$\sigma$ range in $\alpha$ found by \citet{2001MNRAS.322..231K} \citep[see also][]{2018Sci...359...69S}.
For $\beta$ we use -1.0 (negative values of $\beta$ favor binaries with low $q_{\rm ZAMS}$ or unequal binary component masses) and 0.0.
$\beta= 0.0$ is according \citet{2012Sci...337..444S, 2014ApJS..213...34K} and $\beta= -1.0$ is to allow comparison of our results against those of \citet{2011A&A...528A.131C}.
For $\gamma$ we use 0.0 and -0.22. The former is {\"O}pik's law \citep{1924PTarO..25f...1O}, while the latter is according \citet{2014ApJS..213...34K}. 
Note that, though the distributions of \cite{2012Sci...337..444S} and \cite{2014ApJS..213...34K} are only valid up to $P_{\rm orb, ZAMS}=2000$ and 3300 days, respectively, we assume that they hold up to $\simeq 6310$ days. However, we note that we find good agreement with SN IIb rates computed using the distributions of \citet[][discussed next]{2017ApJS..230...15M} that are valid for larger values of $P_{\rm orb, ZAMS}$.

In addition to the simple distributions discussed above, we also consider recent distributions derived by \citet{2017ApJS..230...15M} to calculate SN IIb rates.
We simulate a large population ($10^6$ samples) of single and binary stars with initial (in binary systems, primary) mass $M_{\rm ZAMS} \geq 8 M_\sun$ using a Monte Carlo technique and match the sampled initial properties to our model grids to compute rates.
For the distribution of \citet{2017ApJS..230...15M} an IMF needs to be assumed. We assume a Salpeter IMF with different values of $\alpha=$ 1.6, 2.3, and 3.0.

As we do not model the evolution of secondaries after the primary undergoes CC, we consider two extreme cases in order to bracket our rates: either all of them explode as SNe of a type other than IIb (which is the lower limit on our predicted SN IIb rates), or that none of them explode as SNe (which is the upper limit on our predicted SN IIb rates). While it is possible that the secondaries explode as SNe IIb, we expect their contribution to be at most equal to that of single stars. After the first SN explosion more than $\sim 80\%$ of systems are expected to become unbound due to the kick on the newly formed neutron star \citep{2018arXiv180409164R}, such that the companion would evolve as an effectively single star. Of the few systems that remain bound, RLO is expected to lead to unstable mass transfer and a merger due to the the extreme mass ratio between the neutron star and the massive donor.

\subsection{Single and Binary SN IIb Rates at Solar and Low Metallicity}

\begin{deluxetable*}{lccccccc} 
\centering
\tabletypesize{\scriptsize}
\tablecaption{{{Theoretical versus observationally inferred SN IIb relative rates}} \label{t:ratesmain}}
\tablewidth{0pt}
\tablehead{
\colhead{} & \multicolumn{4}{c}{Theoretical (this paper)} & \multicolumn{3}{c}{Observational}
\\
\cmidrule(lr){2-5} \cmidrule(lr){6-8}
\colhead{metallicity} & \colhead{single} & \multicolumn{2}{c}{binary} & \colhead{single+binary} & \colhead{\cite{2011MNRAS.412.1473L}} & \colhead{\cite{2011MNRAS.412.1522S}} & \colhead{\cite{2017ApJ...837..121G}} 
}
\startdata
high ($Z_\odot$) 	& $0-3.6$ 	& \multicolumn{2}{c}{$0.1-2.2$}			& $0.1-4.1$ 	& $11.9^{+3.9}_{-3.6}$	& $10.6^{+3.6}_{-3.1}$	& $10^{+3}_{-3}$ \\
\cmidrule{3-4}
				&		& $\epsilon=0.5$ 	& $\epsilon=0.1$	&			&					&					& \\
\cmidrule(lr){3-3} \cmidrule(lr){4-4}
				&		& $0.1-1.2$		& $0.1-2.2$		&			&					&					& \\
\midrule
low ($Z_\odot/4$) 	& $0$ 	& \multicolumn{2}{c}{$0.4-15$} 			& $0.5-15$ 	& \nodata				& \nodata				& $19^{+15}_{-10}$ \\
\cmidrule{3-4}
				&		& $\epsilon=0.5$	& $\epsilon=0.1$	&			&					&					& \\
\cmidrule(lr){3-3} \cmidrule(lr){4-4}
				&		& $0.4-4.7$		& $2-15$			&			&					&					&
\enddata
\tablenotetext{}{{\bf (1)} SN IIb relative rate is defined as the percentage of SNe IIb out of all CC SNe.}
\tablenotetext{}{{\bf (2)} Low metallicity rates are lower limits (see text).}
\tablenotetext{}{{\bf (3)} We quote rates from \citet{2017ApJ...837..121G} using their cut of $3\times 10^9M_\sun$ on SN host galaxy stellar mass, as this cut separates galaxies below and above mass of the LMC.}
\end{deluxetable*}


Table \ref{t:ratesmain} summarizes our results for SNe IIb relative rates.
Tables \ref{t:ratessolar} and \ref{t:rateslow} list single- and binary-star (with $\epsilon$ = 0.5 and 0.1) SN IIb rates at solar and low metallicity, respectively, for all values for $f_{\rm bin}$, $\alpha$, $\beta$, and $\gamma$ and four definitions for SN IIb progenitors (see Section \ref{s:channels}), namely
\begin{enumerate}
\item $0.01M_\sun \leq M_{\rm H~env,preSN(,1)} \leq 1M_\sun$ (`fiducial definition'),
\item $0.01M_\sun \leq M_{\rm H~env,preSN(,1)} \leq 0.5M_\sun$,
\item $0.01M_\sun \leq M_{\rm H~env,preSN(,1)} \leq 1M_\sun$ and $2M_\sun \leq M_{\rm He~core,preSN(,1)} \leq 6M_\sun$, and
\item $0.01M_\sun \leq M_{\rm H~env,preSN(,1)} \leq 0.5M_\sun$ and $2M_\sun \leq M_{\rm He~core,preSN(,1)} \leq 6M_\sun$.
\end{enumerate}
Values shown in the tables correspond to lower limits on SN IIb rates, where we assume that all secondaries explode as SNe of a different type. The upper limit on SN IIb rates, assuming that none of the secondaries explode as SNe, can be obtained by multiplying the rates by $(1+f_{\rm bin})$.

Table \ref{t:ratesMD17} lists single- and binary-star (with $\epsilon$ = 0.5 and 0.1) SN IIb rates at solar and low metallicity computed using the distributions of \citet{2017ApJS..230...15M} for all values of $\alpha$ and all four definitions for SN IIb progenitors.
For these we compute and quote lower and upper limits on the rates, assuming that all or none of the secondaries explode as SNe, respectively.
These rates are typically 2-5 times lower than those from our `fiducial distributions' (using $f_{\rm bin} = 0.88$; computed value for our simulated population).

The computed rates increase by $<0.01\%$ when we require that primary stars transfer at least 0.1\% (instead of 1\%) of their initial mass in RLO to qualify as a `binary'. 
The low-metallicity rates are more strongly robust to this definition; there are relatively few mildly interacting binaries at low metallicity.
As mentioned in Section \ref{ss:low}, we limit the initial primary mass to $M_{\rm ZAMS,1} \simeq 25M_\sun$ in our low-metallicity models even though the corresponding single-star models retain large envelopes.
If we assume that all binaries in the unmodeled parameter space lead to SNe IIb (i.e. $M_{\rm ZAMS,1} \gtrsim 25M_\sun$, $q_{\rm ZAMS}$ = 0.4 -- 1.0, and $\logten(P_{\rm orb, ZAMS}/$d$)$ = 3.0 -- 3.7), they would contribute at most $2.7\%$ in rates for the most favorable priors and assuming that no secondaries lead to SNe. 

In Table \ref{t:ratesmain} we report conservative ranges for theoretical SNe IIb relative rates. The lower limits are the smallest values from Tables \ref{t:ratessolar} and \ref{t:rateslow} and assuming that all secondaries explode as SNe of a type other than IIb, while the upper limits are the largest values from those tables and assuming that no secondaries explode as SNe.
Therefore values in Table 2 span the range of theoretical SN IIb rates obtained by varying prior distributions of single- and binary-star birth properties and structural criteria for defining SN IIb progenitors.

We note that our binary SN IIb rates at low metallicity represent lower limits as we do not compute progenitors arising via case A mass transfer.
We do not expect SN IIb progenitors via case A mass transfer at solar metallicity because their residual hydrogen envelope after the mass transfer phase will be roughly as massive as that for binary stars that initiate mass transfer on the HG \citep{2017A&A...608A..11G} and will therefore also be stripped by winds before CC. 

Overall, our model (single and binary) SNe IIb contribute to $0-4.1\%$ and $0.4-15\%$ of all CC SNe at solar and low metallicity, respectively. 
SN IIb rates from observations is $\sim 11\%$ \citep{2011MNRAS.412.1473L,2011MNRAS.412.1522S,2017ApJ...837..121G} at high metallicity. 
Our solar-metallicity models can account for less than half the observationally inferred rate at high metallicity.
On the other hand, there is evidence that SN IIb rates may be higher in galaxies less massive than the LMC \citep{2010ApJ...721..777A,2017ApJ...837..121G}. Note that observationally inferred rates as a function of galaxy mass/metallicity should be interpreted with caution as they are affected by small sample sizes.
However, if this trend holds up in future investigations using large sample sizes, our models can account for the implied SNe IIb rates at low metallicity. 
However, this is only true for models with low mass transfer efficiency.

Applying the most restrictive criteria for defining SN IIb progenitors [$M_{\rm H~env,preSN(,1)} \leq 0.5M_\sun$ and $2M_\sun \leq M_{\rm He~core,preSN(,1)} \leq 6M_\sun$], makes the disagreement between theoretical ($0.1 - 1.4$\%) and observed rates at solar metallicity worse.  
At low metallicity, however, though theoretical rates ($0.5-12$\%) also decrease, our most optimistic rates are still consistent with observed rates.
An analogous result was also found by \citet{2016MNRAS.457..328L}, who found that low-metallicity binaries with primary masses $8-20M\sun$ can explain the observed distribution of SN IIb (and also SE SN generally) ejecta masses.

\pagebreak
\section{Discussions and Conclusions} \label{s:conclusions}

\subsection{Summary of Results}

In this paper, we use non-rotating single- and binary-star models at solar and low metallicities covering a broad parameter space (see Table \ref{t:pspace}) to investigate evolutionary channels and parameter space for progenitors of SNe IIb. 
We find that metallicity and mass transfer efficiency play important roles in shaping the parameter space towards SNe IIb.
Both of these effect the evolution of case EB binaries (binaries that initiate MT on the HG), where lower metallicities and mass transfer efficiencies are favorable for the production of SNe IIb. 
Since the range of binary configurations that lead to case EB mass transfer is very wide, these binaries have a strong effect in shaping the parameter space, and therefore rates for SNe IIb.

We use our models to compute theoretical SN IIb rates and compare them to observationally constrained values. 
We attempt to account for the uncertainties in the prior distributions of single- and binary-star birth properties and criteria used to define SN IIb progenitor structures by spanning possible values for them. 
This allows us to investigate to what extent our models are a priori able to explain observed SN IIb rates.

We find that solar-metallicity single- and binary-star models, by themselves or together, account for less than half the observed SN IIb rate at solar metallicity. 
Our most optimistic rate is inconsistent within $2\sigma$ of the observationally inferred rates. 
Also, at high metallicity, singles and binaries contribute to roughly equal numbers (singles slightly more) of SNe IIb.
At low metallicity, however, our models can account for the implied rate.
Here, binaries are the sole producers of SNe IIb. The group of case EB binary SNe IIb (which in turn is courtesy of lower wind mass-loss rates at low metallicity) is responsible for this change.
Further, even using our most restrictive criterion on a SN IIb progenitor structure, theoretical rates at low metallicity are still consistent with observed rates within the reported uncertainities (though these are quite large).

If wind mass-loss rates are lower than those used in our models (which may be the case as discussed in Section \ref{ss:physics}), it might be instructive to compare observed rates at high metallicity to our low-metallicity models \citep{2017ApJ...840...10Y}. We find that rates for our low-metallicity and low mass transfer efficiency models are consistent with observed rates at high metallicity. 
This is true even after using our most restrictive definition of a SN IIb progenitor. 
However, this requires $f_{\rm bin} > 0.5$ and priors favoring equal mass binaries and favoring a flat orbital period distribution to be consistent within reported uncertainties for observationally inferred rates.
We discuss implications of this in Section \ref{ss:takeaways}.

\subsection{Implications of Assumed Physics} \label{ss:physics}

Although we have tried to be conservative in our predicted upper and lower limits for rates, it is instructive to point out additional physical uncertainties in our models.

Wind mass-loss rates have an important effect on single- and binary-star channels towards SNe IIb. 
Stellar winds in our \mesa\ models are computed using different wind mass-loss prescriptions at various evolutionary phases. 
WR mass-loss rates have the strongest impact on evolutionary outcomes. However, these rates are also known to be quite uncertain, particularly for low mass WR stars. Most estimates, including those used in this work are expected to be overestimated due to clumping in the winds \citep{2014ARA&A..52..487S}. 
Lower wind mass-loss rates would expand the parameter space for binary SNe IIb \citep{2017ApJ...840...10Y,2019MNRAS.486.4451G}. 
In addition, though it is commonly assumed that mass-loss rates from red supergiants decrease with metallicity, some studies suggest that they are more or less metallicity independent \citep{2000A&A...354..125V, 2005A&A...438..273V, 2017MNRAS.465..403G}. If this is the case, we would expect low-metallicity single stars to produce SN IIb at a relative rate similar to that of high-metallicity single stars.

Another important point is that we have not included stellar rotation. Rotation is thought to play an important role in massive star evolution, with very rapidly rotating stars even being predicted to evolve chemically homogeneously due to rotationally enhanced mixing \citep{1987A&A...178..159M}. Rotationally enhanced mass-loss rates can also reduce the minimum mass required for a single star to remove its hydrogen envelope \citep{2003A&A...404..975M}, thus increasing the rate of SN IIb progenitors from single stars. 
However, \citet{2013A&A...550L...7G} found that the initial mass range that produces SNe IIb does not change much for their rotating models with $v_{rot} \sim 250$ km s$^{-1}$. 
Further, observations show that the bulk of the massive star population rotates at lower velocities \citep{2013A&A...560A..29R,2015A&A...580A..92R}, and that stars rotating at faster rates can potentially be accounted for in terms of spun-up secondaries in post-mass transfer systems \citep{2013ApJ...764..166D}.
Properly assessing the impact of rotation, along with the effect of tides and eccentricity (which is likely small due to the wide orbits considered here) would require future calculations that include these effects, but they are beyond the scope of the present study.



\subsection{Key Takeaways and Future Directions} \label{ss:takeaways}

Our results have the following implications for progenitor channels to SNe IIb. In order to account for the observationally inferred SN IIb rates,
\begin{enumerate}
\item solar-metallicity wind mass-loss rates need to be lower than those used in our models: lower wind mass-loss rates increase the contribution of binaries to SNe IIb, by populating the parameter space available via case EB mass transfer, while driving down the contribution from single stars.
\item low mass transfer efficiencies are needed: this allows case EB accretors to thermally adjust to the transferred mass and avoid contact.
\end{enumerate}
We note that it is likely that even after accounting for robust mass-loss rates, the uncertainty in prior distributions of stellar birth properties and SN IIb definitions will still pose significant challenges for comparing theoretical models to observations.
Therefore, in order to address the question of SNe IIb progrenitors we need four pieces of information: 
(1) accurate determinations of mass-loss rates for stripped stars as a function of metallicity, (2) constraints on structural properties of SN IIb progenitor stars, (3) robust distributions for single- and binary-star birth properties, and (4) on the observational side, SN IIb rates as a function of metallicity.
An exploration of case A mass transfer towards SNe IIb at low metallicity and merger channels would also be worthwhile.


In this paper we focussed on understanding the evolutionary channels and parameter space of SN IIb progenitors. In a follow-up paper we will focus on using our models to draw statistical comparisons to all three independent observational probes into SN IIb progenitors: (1) direct progenitor detections in archival images, (2) progenitor structure and explosion parameters from analyzing multi-band light-curves, and (3) circumstellar medium properties from X-ray/radio observations.
The aim would be to confirm that our models are able to reproduce the full range of observational constraints for SNe IIb, that the results of such comparison is consistent with conclusions in this paper, and finally, to identify regions in the observational parameter space that are implied from our models but remain unprobed by observations to help guide observing strategies.

The advent of high-cadence surveys like the ZTF \citep{2014htu..conf...27B, 2017NatAs...1E..71B}, ASSA-SN \citep{2017PASP..129j4502K}, DLT40 \citep{2017ApJ...848L..24V}, KAIT \citep{2001ASPC..246..121F}, and, at the turn of the decade, LSST \citep{2002SPIE.4836...10T,2008arXiv0805.2366I} 
have created the need for a comprehensive database of theoretical models to provide reliable progenitor characterization and in the case of binary progenitors, their companions. Conversely, the wealth of data that will be available will provide unprecedented statistics on progenitor properties, along with their dependence on properties such as host galaxy metallicity. 
Efforts such as the one undertaken here are therefore imperative in order to understand the physics of massive star evolution.

\acknowledgments
NS, PM, and VK acknowledge support from NSF grant AST-1517753. NS acknowledges support from NSF grant DGE-1450006 and from Northwestern University. 
VK also acknowledges support by the Canadian Institute for Advanced Research as a Fellow.
This work was performed in part at the Aspen Center for Physics, which is supported by National Science Foundation through grant PHY-1607611.
We would also like to thank Mads S{\o}rensen for providing a Python version of Maxwell Moe's script to simulate a population of stars from distributions of \citet{2017ApJS..230...15M}.
This research was supported by computational resources provided for the Quest high performance computing facility at Northwestern University which is jointly supported by the National Science Foundation through grant NSF PHY-1126812, the Office of the Provost, the Office for Research, and Northwestern University Information Technology.





\appendix

\section{Numerical Tests} \label{s:conv}

\begin{figure*}
\begin{center}
\includegraphics[width=\textwidth]{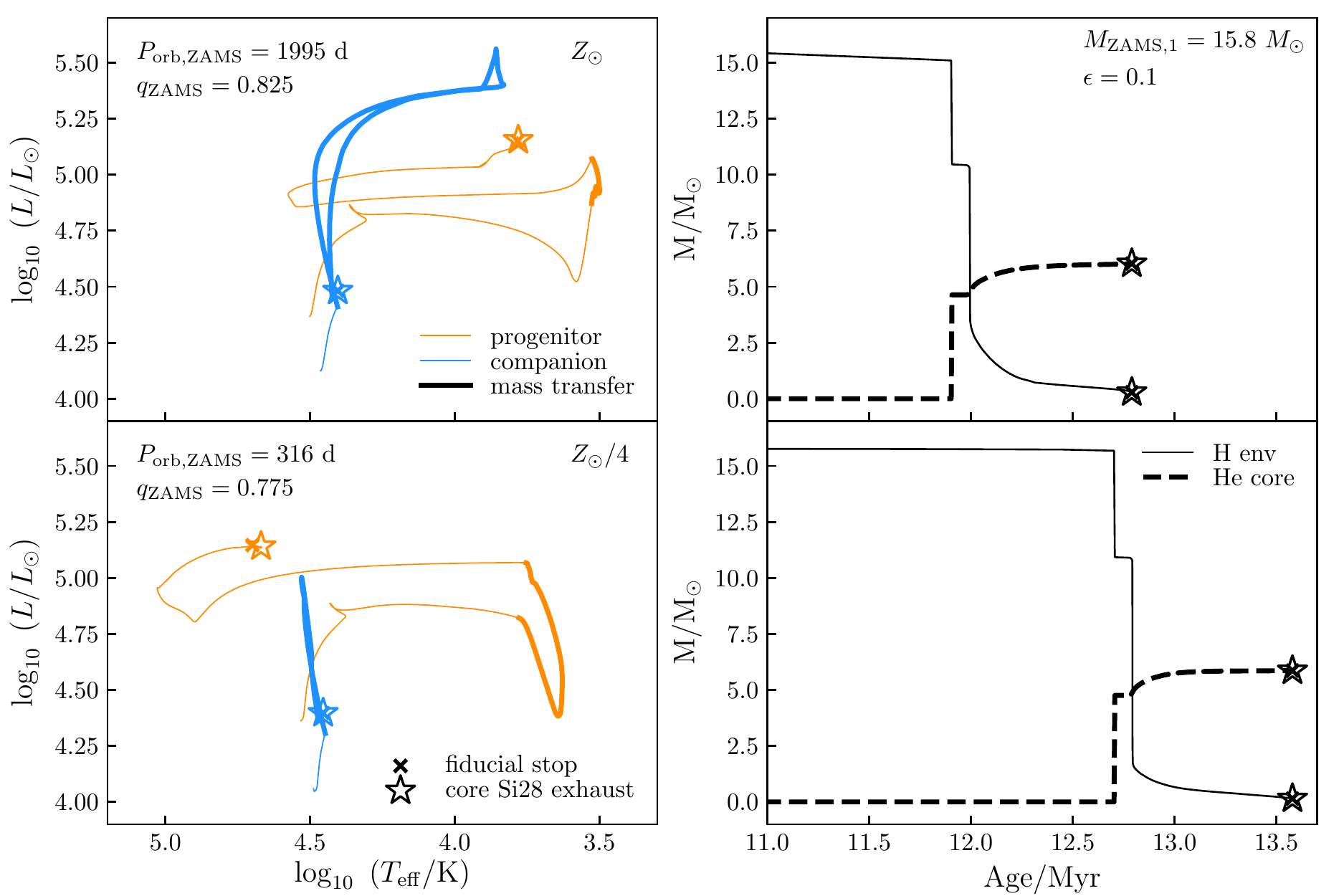}
\end{center}
\caption[]{{\it Left}: H-R diagrams showing evolution of representative solar- (top) and low- (bottom) metallicity binary-star models to core silicon exhaustion. Line colors and weights have the same meaning as in Figure \ref{f:binary}. 
{\it Right}: Evolution of primary hydrogen envelope (solid line) and helium core mass (dashed line) as a function of binary age for the corresponding models on the left.
In all panels, crosses (stars) denote the point where carbon (silicon) is exhausted in the core of the primary.
The properties of all models have converged to those very near CC by the fiducial evolutionary stopping criterion used in this work.
\label{f:convergence}}
\end{figure*}

We stop the evolution of our single- and binary-star models close to core carbon depletion (according to the definitions in Section \ref{s:models}) and assume that their global characteristics of interest when investigating SN IIb progenitors are roughly the same at CC. To confirm that this is indeed the case, we run representative binary-star models at solar (case LB mass transfer type) 
and low (case EB mass transfer type, also investigated in Section \ref{s:channels} and Figure \ref{f:gapHR}) metallicity until central silicon mass fraction in the primary drops below $10^{-6}$ (at which point the models are a few hours from CC). We then check whether the primary hydrogen envelope and helium core mass and H-R locations of both components differ significantly at this later evolutionary stage. 

Figure \ref{f:convergence} shows the result of this test. We find that both models have converged by the fiducial evolutionary stopping criterion used in this work to that at the more advanced evolutionary stage for all properties relevant to our investigation.


\begin{deluxetable*}{cccccccccccccccc} 
\centering
\tabletypesize{\scriptsize}
\tablecaption{{{SN IIb relative rates at solar metallicity}} \label{t:ratessolar}}
\tablewidth{0pt}
\tablehead{
\colhead{} & \colhead{} & \colhead{} & \colhead{} & \multicolumn{6}{c}{$M_{\rm He~core} \geq 0M_\sun$}  & \multicolumn{6}{c}{$2M_\sun \leq M_{\rm He~core} \leq 6M_\sun$} 
\\
\cmidrule(lr){5-10} \cmidrule(lr){11-16} 
\colhead{} & \colhead{} & \colhead{} & \colhead{} & \multicolumn{3}{c}{$M_{\rm H~env} \leq 1.0M_\sun$} & \multicolumn{3}{c}{$M_{\rm H~env} \leq 0.5M_\sun$} & \multicolumn{3}{c}{$M_{\rm H~env} \leq 1.0M_\sun$} & \multicolumn{3}{c}{$M_{\rm H~env} \leq 0.5M_\sun$} 
\\
\cmidrule(lr){5-7} \cmidrule(lr){8-10} \cmidrule(lr){11-13} \cmidrule(lr){14-16} 
\colhead{$f_{\rm bin}$} & \colhead{$\alpha$} & \colhead{$\beta$} & \colhead{$\gamma$} & \colhead{single} & \multicolumn{2}{c}{binary} & \colhead{single} & \multicolumn{2}{c}{binary} & \colhead{single} & \multicolumn{2}{c}{binary} & \colhead{single} & \multicolumn{2}{c}{binary}
\\ 
\cmidrule(lr){6-7} \cmidrule(lr){9-10} \cmidrule(lr){12-13} \cmidrule(lr){15-16} 
\colhead{} & \colhead{} & \colhead{} & \colhead{} & \colhead{} & \colhead{$\epsilon = 0.5$} & \colhead{$\epsilon = 0.1$} & \colhead{} & \colhead{$\epsilon = 0.5$} & \colhead{$\epsilon = 0.1$} & \colhead{} & \colhead{$\epsilon = 0.5$} & \colhead{$\epsilon = 0.1$} & \colhead{} & \colhead{$\epsilon = 0.5$} & \colhead{$\epsilon = 0.1$}
}
\startdata
			 	& 				& \multirow{2}{*}{-1.0}	& 0.0 	 &2.8 & 0.14 & 0.21 & 1.9 & 0.11 & 0.16 & 0.0 & 0.12 & 0.17 & 0.0 & 0.08 & 0.13\\
 				& \multirow{2}{*}{1.6}		& 					& -0.22 	 &2.8 & 0.12 & 0.18 & 1.9 & 0.09 & 0.14 & 0.0 & 0.1 & 0.14 & 0.0 & 0.07 & 0.11\\
								\cline{3-16}

 				& 				& \multirow{2}{*}{0.0}		& 0.0 	 &2.8 & 0.2 & 0.33 & 1.9 & 0.14 & 0.26 & 0.0 & 0.16 & 0.26 & 0.0 & 0.1 & 0.2\\
 				& 				& 					& -0.22 	 &2.8 & 0.17 & 0.28 & 1.9 & 0.12 & 0.22 & 0.0 & 0.13 & 0.22 & 0.0 & 0.09 & 0.17\\
				\cline{2-16}

 				&				& \multirow{2}{*}{-1.0}	& 0.0 	 &2.8 & 0.22 & 0.31 & 1.9 & 0.16 & 0.24 & 0.0 & 0.18 & 0.26 & 0.0 & 0.13 & 0.2\\
\multirow{2}{*}{0.25} 	& \multirow{2}{*}{2.3}		& 	& -0.22 	 &2.8 & 0.18 & 0.27 & 1.9 & 0.13 & 0.21 & 0.0 & 0.15 & 0.23 & 0.0 & 0.11 & 0.17\\
								\cline{3-16}

 				&				& \multirow{2}{*}{0.0}		& 0.0 	 &2.8 & 0.29 & 0.49 & 1.9 & 0.2 & 0.38 & 0.0 & 0.24 & 0.41 & 0.0 & 0.16 & 0.31\\
 				&				& 					& -0.22 	 &2.8 & 0.25 & 0.42 & 1.9 & 0.17 & 0.32 & 0.0 & 0.2 & 0.35 & 0.0 & 0.13 & 0.27\\
				\cline{2-16}

 				&				& \multirow{2}{*}{-1.0}	& 0.0 	 &2.0 & 0.23 & 0.34 & 1.3 & 0.17 & 0.26 & 0.0 & 0.21 & 0.3 & 0.0 & 0.14 & 0.22\\
 				& \multirow{2}{*}{3.0}		& 					& -0.22 	 &2.0 & 0.2 & 0.29 & 1.3 & 0.14 & 0.22 & 0.0 & 0.17 & 0.26 & 0.0 & 0.12 & 0.19\\
								\cline{3-16}

 				&				& \multirow{2}{*}{0.0}		& 0.0 	 &2.0 & 0.32 & 0.53 & 1.3 & 0.22 & 0.41 & 0.0 & 0.27 & 0.46 & 0.0 & 0.18 & 0.35\\
 				&				& 					& -0.22 	 &2.0 & 0.27 & 0.46 & 1.3 & 0.18 & 0.35 & 0.0 & 0.23 & 0.4 & 0.0 & 0.15 & 0.3\\
\hline

			 	&				& \multirow{2}{*}{-1.0}	& 0.0 	 &1.6 & 0.24 & 0.35 & 1.1 & 0.18 & 0.27 & 0.0 & 0.2 & 0.28 & 0.0 & 0.14 & 0.21\\
			 	& \multirow{2}{*}{1.6}		& 					& -0.22 	 &1.6 & 0.2 & 0.3 & 1.1 & 0.15 & 0.23 & 0.0 & 0.17 & 0.24 & 0.0 & 0.12 & 0.18\\
								\cline{3-16}

			 	&				& \multirow{2}{*}{0.0}		& 0.0 	 &1.6 & 0.33 & 0.55 & 1.1 & 0.24 & 0.43 & 0.0 & 0.26 & 0.44 & 0.0 & 0.17 & 0.33\\
			 	&				& 					& -0.22 	 &1.6 & 0.28 & 0.47 & 1.1 & 0.2 & 0.37 & 0.0 & 0.22 & 0.37 & 0.0 & 0.15 & 0.29\\
				\cline{2-16}

			 	&				& \multirow{2}{*}{-1.0}	& 0.0 	 &1.6 & 0.36 & 0.52 & 1.0 & 0.26 & 0.4 & 0.0 & 0.3 & 0.44 & 0.0 & 0.21 & 0.33\\
\multirow{2}{*}{0.50}	& \multirow{2}{*}{2.3}		& 	& -0.22 	 	& 1.6 	& 0.3 & 0.45 & 1.0 & 0.22 & 0.34 & 0.0 & 0.26 & 0.38 & 0.0 & 0.18 & 0.28\\
								\cline{3-16}

				&				& \multirow{2}{*}{0.0}		& 0.0 	 &1.6 & 0.49 & 0.81 & 1.0 & 0.34 & 0.63 & 0.0 & 0.4 & 0.68 & 0.0 & 0.26 & 0.51\\
				&				& 					& -0.22 	 &1.6 & 0.41 & 0.7 & 1.0 & 0.29 & 0.54 & 0.0 & 0.34 & 0.59 & 0.0 & 0.22 & 0.44\\
				\cline{2-16}

				&				& \multirow{2}{*}{-1.0}	& 0.0 	 &1.1 & 0.39 & 0.57 & 0.74 & 0.28 & 0.43 & 0.0 & 0.34 & 0.5 & 0.0 & 0.24 & 0.37\\
			 	& \multirow{2}{*}{3.0}		& 					& -0.22 	 &1.1 & 0.33 & 0.49 & 0.74 & 0.24 & 0.37 & 0.0 & 0.29 & 0.43 & 0.0 & 0.2 & 0.32\\
								\cline{3-16}

			 	&				& \multirow{2}{*}{0.0}		& 0.0 	 &1.1 & 0.53 & 0.89 & 0.74 & 0.36 & 0.68 & 0.0 & 0.45 & 0.77 & 0.0 & 0.29 & 0.58\\
			 	&				& 					& -0.22 	 &1.1 & 0.45 & 0.76 & 0.74 & 0.31 & 0.59 & 0.0 & 0.38 & 0.67 & 0.0 & 0.25 & 0.5\\
\hline

			 	&				& \multirow{2}{*}{-1.0}	& 0.0 	 &1.0 & 0.29 & 0.41 & 0.68 & 0.21 & 0.32 & 0.0 & 0.23 & 0.33 & 0.0 & 0.16 & 0.25\\
			 	& \multirow{2}{*}{1.6}		& 					& -0.22 	 &1.0 & 0.24 & 0.35 & 0.68 & 0.18 & 0.28 & 0.0 & 0.2 & 0.28 & 0.0 & 0.14 & 0.21\\
								\cline{3-16}

			 	&				& \multirow{2}{*}{0.0}		& 0.0 	 &1.0 & 0.39 & 0.64 & 0.68 & 0.28 & 0.5 & 0.0 & 0.31 & 0.51 & 0.0 & 0.2 & 0.39\\
			 	&				& 					& -0.22 	 &1.0 & 0.33 & 0.55 & 0.68 & 0.24 & 0.43 & 0.0 & 0.26 & 0.44 & 0.0 & 0.17 & 0.34\\
				\cline{2-16}

			 	&				& \multirow{2}{*}{-1.0}	& 0.0 	 &1.0 & 0.42 & 0.62 & 0.66 & 0.31 & 0.47 & 0.0 & 0.36 & 0.52 & 0.0 & 0.25 & 0.39\\
\multirow{2}{*}{0.65} 	& \multirow{2}{*}{2.3}		& 	& -0.22 	 &1.0 & 0.36 & 0.53 & 0.66 & 0.26 & 0.41 & 0.0 & 0.3 & 0.44 & 0.0 & 0.22 & 0.33\\
								\cline{3-16}

				&				& \multirow{2}{*}{0.0}		& 0.0 	 &1.0 & 0.58 & 0.96 & 0.66 & 0.4 & 0.74 & 0.0 & 0.47 & 0.81 & 0.0 & 0.31 & 0.61\\
			 	&				& 					& -0.22 	 &1.0 & 0.49 & 0.83 & 0.66 & 0.34 & 0.64 & 0.0 & 0.4 & 0.69 & 0.0 & 0.27 & 0.52\\
				\cline{2-16}

			 	&				& \multirow{2}{*}{-1.0}	& 0.0 	 &0.72 & 0.46 & 0.67 & 0.47 & 0.33 & 0.51 & 0.0 & 0.4 & 0.59 & 0.0 & 0.28 & 0.43\\
			 	& \multirow{2}{*}{3.0}		& 					& -0.22 	 &0.72 & 0.39 & 0.58 & 0.47 & 0.28 & 0.44 & 0.0 & 0.34 & 0.5 & 0.0 & 0.24 & 0.38\\
								\cline{3-16}

			 	&				& \multirow{2}{*}{0.0}		& 0.0 	 &0.72 & 0.62 & 1.1 & 0.47 & 0.43 & 0.8 & 0.0 & 0.53 & 0.91 & 0.0 & 0.35 & 0.69\\
			 	&				& 					& -0.22 	 &0.72 & 0.53 & 0.9 & 0.47 & 0.36 & 0.69 & 0.0 & 0.45 & 0.79 & 0.0 & 0.29 & 0.59\\
\hline

			 	&				& \multirow{2}{*}{-1.0}	& 0.0 	 &0.53 & 0.32 & 0.47 & 0.36 & 0.24 & 0.36 & 0.0 & 0.26 & 0.37 & 0.0 & 0.19 & 0.28\\
			 	& \multirow{2}{*}{1.6}		& 					& -0.22 	 &0.53 & 0.27 & 0.4 & 0.36 & 0.2 & 0.31 & 0.0 & 0.22 & 0.32 & 0.0 & 0.16 & 0.24\\
								\cline{3-16}

			 	&				& \multirow{2}{*}{0.0}		& 0.0 	 &0.53 & 0.44 & 0.73 & 0.36 & 0.31 & 0.57 & 0.0 & 0.34 & 0.58 & 0.0 & 0.23 & 0.44\\
			 	&				& 					& -0.22 	 &0.53 & 0.37 & 0.62 & 0.36 & 0.27 & 0.49 & 0.0 & 0.29 & 0.5 & 0.0 & 0.2 & 0.38\\
				\cline{2-16}

			 	&				& \multirow{2}{*}{-1.0}	& 0.0 	 &0.52 & 0.48 & 0.69 & 0.35 & 0.35 & 0.53 & 0.0 & 0.41 & 0.58 & 0.0 & 0.29 & 0.44\\
\multirow{2}{*}{0.80}	& \multirow{2}{*}{2.3}		& 	& -0.22 	 &0.52 & 0.41 & 0.59 & 0.35 & 0.3 & 0.46 & 0.0 & 0.34 & 0.5 & 0.0 & 0.24 & 0.38\\
								\cline{3-16}

				&				& \multirow{2}{*}{0.0}		& 0.0 	 &0.52 & 0.65 & 1.1 & 0.35 & 0.45 & 0.84 & 0.0 & 0.53 & 0.91 & 0.0 & 0.35 & 0.69\\
			 	&				& 					& -0.22 	 &0.52 & 0.55 & 0.93 & 0.35 & 0.39 & 0.72 & 0.0 & 0.45 & 0.78 & 0.0 & 0.3 & 0.59\\
				\cline{2-16}

			 	&				& \multirow{2}{*}{-1.0}	& 0.0 	 &0.38 & 0.52 & 0.76 & 0.25 & 0.38 & 0.58 & 0.0 & 0.46 & 0.66 & 0.0 & 0.32 & 0.49\\
			 	& \multirow{2}{*}{3.0}		& 					& -0.22 	 &0.38 & 0.44 & 0.65 & 0.25 & 0.32 & 0.5 & 0.0 & 0.39 & 0.57 & 0.0 & 0.27 & 0.42\\
								\cline{3-16}

			 	&				& \multirow{2}{*}{0.0}		& 0.0 	 &0.38 & 0.7 & 1.2 & 0.25 & 0.48 & 0.91 & 0.0 & 0.6 & 1.0 & 0.0 & 0.39 & 0.77\\
			 	&				& 					& -0.22 	 &0.38 & 0.59 & 1.0 & 0.25 & 0.41 & 0.78 & 0.0 & 0.51 & 0.89 & 0.0 & 0.33 & 0.67
\enddata
\tablenotetext{}{{\bf (1)} $M_{\rm He~core} \equiv M_{\rm He~core,preSN(,1)}$: Residual hydrogen envelope mass in the progenitor at CC\\
{\bf (2)} $M_{\rm H~env} \equiv M_{\rm H~env,preSN(,1)}$ : Helium core mass of the progenitor at CC} 
\tablecomments{$f_{\rm bin}$ is fraction of binary systems and $\alpha$, $\beta$, and $\gamma$ are parameters for the priors on the initial mass, $\logten M_{\rm ZAMS}$, initial mass ratio, $q_{\rm ZAMS}$, and initial orbital period, $P_{\rm orb}$, respectively (see Eqs. \ref{e:IMF}, \ref{e:f(q)}, and \ref{e:f(P)}). SN IIb rate is defined as fraction of SNe IIb vs all CC SNe percent. 
}
\end{deluxetable*}

\begin{deluxetable*}{cccccccccccccccc} 
\centering
\tabletypesize{\scriptsize}
\tablecaption{{{SN IIb relative rates at low metallicity}} \label{t:rateslow}}
\tablewidth{0pt}
\tablehead{
\colhead{} & \colhead{} & \colhead{} & \colhead{} & \multicolumn{6}{c}{$M_{\rm He~core} \geq 0M_\sun$}  & \multicolumn{6}{c}{$2M_\sun \leq M_{\rm He~core} \leq 6M_\sun$} 
\\
\cmidrule(lr){5-10} \cmidrule(lr){11-16} 
\colhead{} & \colhead{} & \colhead{} & \colhead{} & \multicolumn{3}{c}{$M_{\rm H~env} \leq 1.0M_\sun$} & \multicolumn{3}{c}{$M_{\rm H~env} \leq 0.5M_\sun$} & \multicolumn{3}{c}{$M_{\rm H~env} \leq 1.0M_\sun$} & \multicolumn{3}{c}{$M_{\rm H~env} \leq 0.5M_\sun$} 
\\
\cmidrule(lr){5-7} \cmidrule(lr){8-10} \cmidrule(lr){11-13} \cmidrule(lr){14-16} 
\colhead{$f_{\rm bin}$} & \colhead{$\alpha$} & \colhead{$\beta$} & \colhead{$\gamma$} & \colhead{single} & \multicolumn{2}{c}{binary} & \colhead{single} & \multicolumn{2}{c}{binary} & \colhead{single} & \multicolumn{2}{c}{binary} & \colhead{single} & \multicolumn{2}{c}{binary}
\\ 
\cmidrule(lr){6-7} \cmidrule(lr){9-10} \cmidrule(lr){12-13} \cmidrule(lr){15-16} 
\colhead{} & \colhead{} & \colhead{} & \colhead{} & \colhead{} & \colhead{$\epsilon = 0.5$} & \colhead{$\epsilon = 0.1$} & \colhead{} & \colhead{$\epsilon = 0.5$} & \colhead{$\epsilon = 0.1$} & \colhead{} & \colhead{$\epsilon = 0.5$} & \colhead{$\epsilon = 0.1$} & \colhead{} & \colhead{$\epsilon = 0.5$} & \colhead{$\epsilon = 0.1$}
}
\startdata
			 	&					& \multirow{2}{*}{-1.0}	& 0.0    &  0.0  &  0.65  &  2.7  &  0.0  &  0.54  &  2.5  &  0.0  &  0.51  &  2.1  &  0.0  &  0.46  &  2.0\\
			 	& \multirow{2}{*}{1.6}		& 					& -0.22  &  0.0  &  0.64  &  2.5  &  0.0  &  0.54  &  2.4  &  0.0  &  0.51  &  2.1  &  0.0  &  0.48  &  2.0\\
									\cline{3-16}

				&					& \multirow{2}{*}{0.0}		& 0.0    &  0.0  &  1.0  &  3.4  &  0.0  &  0.89  &  3.2  &  0.0  &  0.83  &  2.7  &  0.0  &  0.76  &  2.6\\
				&					& 					& -0.22  &  0.0  &  1.0  &  3.3  &  0.0  &  0.91  &  3.1  &  0.0  &  0.83  &  2.6  &  0.0  &  0.78  &  2.5\\
				\cline{2-16}

				&					& \multirow{2}{*}{-1.0}	& 0.0    &  0.0  &  0.71  &  2.9  &  0.0  &  0.6  &  2.7  &  0.0  &  0.59  &  2.4  &  0.0  &  0.54  &  2.3\\
\multirow{2}{*}{0.25} 	& \multirow{2}{*}{2.3}		& 					& -0.22  &  0.0  &  0.71  &  2.8  &  0.0  &  0.62  &  2.6  &  0.0  &  0.59  &  2.3  &  0.0  &  0.55  &  2.2\\
									\cline{3-16}

				&					& \multirow{2}{*}{0.0}		& 0.0    &  0.0  &  1.1  &  3.7  &  0.0  &  1.0  &  3.4  &  0.0  &  0.96  &  3.1  &  0.0  &  0.9  &  2.9\\
				&					& 					& -0.22  &  0.0  &  1.1  &  3.6  &  0.0  &  1.0  &  3.3  &  0.0  &  0.97  &  3.0  &  0.0  &  0.91  &  2.8\\
				\cline{2-16}

				&					& \multirow{2}{*}{-1.0}	& 0.0    &  0.0  &  0.73  &  2.9  &  0.0  &  0.64  &  2.7  &  0.0  &  0.64  &  2.5  &  0.0  &  0.58  &  2.4\\
				& \multirow{2}{*}{3.0}		& 					& -0.22  &  0.0  &  0.73  &  2.8  &  0.0  &  0.65  &  2.7  &  0.0  &  0.64  &  2.5  &  0.0  &  0.59  &  2.4\\
									\cline{3-16}

				&					& \multirow{2}{*}{0.0}		& 0.0    &  0.0  &  1.2  &  3.7  &  0.0  &  1.1  &  3.5  &  0.0  &  1.0  &  3.2  &  0.0  &  0.97  &  3.0\\
				&					& 					& -0.22  &  0.0  &  1.2  &  3.6  &  0.0  &  1.1  &  3.4  &  0.0  &  1.0  &  3.1  &  0.0  &  0.99  &  3.0\\
\hline

				&					& \multirow{2}{*}{-1.0}	& 0.0    &  0.0  &  1.1  &  4.4  &  0.0  &  0.91  &  4.1  &  0.0  &  0.84  &  3.6  &  0.0  &  0.78  &  3.4\\
				& \multirow{2}{*}{1.6}		& 					& -0.22  &  0.0  &  1.1  &  4.3  &  0.0  &  0.91  &  4.0  &  0.0  &  0.86  &  3.5  &  0.0  &  0.8  &  3.3\\
									\cline{3-16}

				&					& \multirow{2}{*}{0.0}		& 0.0    &  0.0  &  1.7  &  5.7  &  0.0  &  1.5  &  5.3  &  0.0  &  1.4  &  4.5  &  0.0  &  1.3  &  4.3\\
				&					& 					& -0.22  &  0.0  &  1.7  &  5.5  &  0.0  &  1.5  &  5.1  &  0.0  &  1.4  &  4.4  &  0.0  &  1.3  &  4.2\\
				\cline{2-16}

				&					& \multirow{2}{*}{-1.0}	& 0.0    &  0.0  &  1.2  &  4.8  &  0.0  &  1.0  &  4.5  &  0.0  &  0.98  &  4.0  &  0.0  &  0.9  &  3.8\\
\multirow{2}{*}{0.50} 	& \multirow{2}{*}{2.3}		& 					& -0.22  &  0.0  &  1.2  &  4.6  &  0.0  &  1.0  &  4.4  &  0.0  &  0.99  &  3.9  &  0.0  &  0.92  &  3.8\\
									\cline{3-16}

				&					& \multirow{2}{*}{0.0}		& 0.0    &  0.0  &  1.9  &  6.2  &  0.0  &  1.7  &  5.7  &  0.0  &  1.6  &  5.1  &  0.0  &  1.5  &  4.8\\
				&					& 					& -0.22  &  0.0  &  1.9  &  5.9  &  0.0  &  1.7  &  5.5  &  0.0  &  1.6  &  5.0  &  0.0  &  1.5  &  4.7\\
				\cline{2-16}

				&					& \multirow{2}{*}{-1.0}	& 0.0    &  0.0  &  1.2  &  4.8  &  0.0  &  1.1  &  4.6  &  0.0  &  1.1  &  4.2  &  0.0  &  0.98  &  4.0\\
				& \multirow{2}{*}{3.0}		& 					& -0.22  &  0.0  &  1.2  &  4.7  &  0.0  &  1.1  &  4.4  &  0.0  &  1.1  &  4.1  &  0.0  &  1.0  &  3.9\\
									\cline{3-16}

				&					& \multirow{2}{*}{0.0}		& 0.0    &  0.0  &  1.9  &  6.2  &  0.0  &  1.8  &  5.8  &  0.0  &  1.7  &  5.4  &  0.0  &  1.6  &  5.1\\
				&					& 					& -0.22  &  0.0  &  1.9  &  6.0  &  0.0  &  1.8  &  5.6  &  0.0  &  1.7  &  5.2  &  0.0  &  1.6  &  4.9\\
\hline

				&					& \multirow{2}{*}{-1.0}	& 0.0    &  0.0  &  1.3  &  5.2  &  0.0  &  1.1  &  4.9  &  0.0  &  1.0  &  4.2  &  0.0  &  0.91  &  4.0\\
				& \multirow{2}{*}{1.6}		& 					& -0.22  &  0.0  &  1.3  &  5.0  &  0.0  &  1.1  &  4.7  &  0.0  &  1.0  &  4.1  &  0.0  &  0.94  &  3.9\\
									\cline{3-16}

				&					& \multirow{2}{*}{0.0}		& 0.0    &  0.0  &  2.0  &  6.8  &  0.0  &  1.8  &  6.2  &  0.0  &  1.6  &  5.4  &  0.0  &  1.5  &  5.0\\
				&					& 					& -0.22  &  0.0  &  2.0  &  6.5  &  0.0  &  1.8  &  6.0  &  0.0  &  1.6  &  5.2  &  0.0  &  1.5  &  4.9\\
				\cline{2-16}

				&					& \multirow{2}{*}{-1.0}	& 0.0    &  0.0  &  1.4  &  5.7  &  0.0  &  1.2  &  5.3  &  0.0  &  1.2  &  4.8  &  0.0  &  1.1  &  4.5\\
\multirow{2}{*}{0.65} 	& \multirow{2}{*}{2.3}		& 					& -0.22  &  0.0  &  1.4  &  5.5  &  0.0  &  1.2  &  5.1  &  0.0  &  1.2  &  4.6  &  0.0  &  1.1  &  4.4\\
									\cline{3-16}

				&					& \multirow{2}{*}{0.0}		& 0.0    &  0.0  &  2.2  &  7.3  &  0.0  &  2.0  &  6.7  &  0.0  &  1.9  &  6.1  &  0.0  &  1.8  &  5.7\\
				&					& 					& -0.22  &  0.0  &  2.2  &  7.0  &  0.0  &  2.0  &  6.5  &  0.0  &  1.9  &  5.9  &  0.0  &  1.8  &  5.6\\
				\cline{2-16}

				&					& \multirow{2}{*}{-1.0}	& 0.0    &  0.0  &  1.4  &  5.7  &  0.0  &  1.3  &  5.4  &  0.0  &  1.3  &  5.0  &  0.0  &  1.2  &  4.8\\
				& \multirow{2}{*}{3.0}		& 					& -0.22  &  0.0  &  1.4  &  5.5  &  0.0  &  1.3  &  5.2  &  0.0  &  1.3  &  4.9  &  0.0  &  1.2  &  4.6\\
									\cline{3-16}

				&					& \multirow{2}{*}{0.0}		& 0.0    &  0.0  &  2.3  &  7.4  &  0.0  &  2.1  &  6.8  &  0.0  &  2.0  &  6.4  &  0.0  &  1.9  &  6.0\\
				&					& 					& -0.22  &  0.0  &  2.3  &  7.1  &  0.0  &  2.1  &  6.6  &  0.0  &  2.0  &  6.2  &  0.0  &  1.9  &  5.8\\
\hline

				&					& \multirow{2}{*}{-1.0}	& 0.0    &  0.0  &  1.4  &  5.9  &  0.0  &  1.2  &  5.5  &  0.0  &  1.1  &  4.7  &  0.0  &  1.0  &  4.5\\
				& \multirow{2}{*}{1.6}		& 					& -0.22  &  0.0  &  1.4  &  5.7  &  0.0  &  1.2  &  5.3  &  0.0  &  1.1  &  4.6  &  0.0  &  1.1  &  4.4\\
									\cline{3-16}

				&					& \multirow{2}{*}{0.0}		& 0.0    &  0.0  &  2.3  &  7.7  &  0.0  &  2.0  &  7.0  &  0.0  &  1.8  &  6.1  &  0.0  &  1.7  &  5.7\\
				&					& 					& -0.22  &  0.0  &  2.3  &  7.3  &  0.0  &  2.0  &  6.8  &  0.0  &  1.8  &  5.9  &  0.0  &  1.8  &  5.6\\
				\cline{2-16}

				&					& \multirow{2}{*}{-1.0}	& 0.0    &  0.0  &  1.6  &  6.4  &  0.0  &  1.4  &  6.0  &  0.0  &  1.3  &  5.4  &  0.0  &  1.2  &  5.1\\
\multirow{2}{*}{0.80} 	& \multirow{2}{*}{2.3}		& 					& -0.22  &  0.0  &  1.6  &  6.2  &  0.0  &  1.4  &  5.8  &  0.0  &  1.3  &  5.2  &  0.0  &  1.2  &  5.0\\
									\cline{3-16}

				&					& \multirow{2}{*}{0.0}		& 0.0    &  0.0  &  2.5  &  8.2  &  0.0  &  2.2  &  7.6  &  0.0  &  2.1  &  6.9  &  0.0  &  2.0  &  6.4\\
				&					& 					& -0.22  &  0.0  &  2.5  &  7.9  &  0.0  &  2.3  &  7.4  &  0.0  &  2.1  &  6.6  &  0.0  &  2.0  &  6.3\\
				\cline{2-16}

				&					& \multirow{2}{*}{-1.0}	& 0.0    &  0.0  &  1.6  &  6.4  &  0.0  &  1.4  &  6.1  &  0.0  &  1.4  &  5.6  &  0.0  &  1.3  &  5.4\\
				& \multirow{2}{*}{3.0}		& 					& -0.22  &  0.0  &  1.6  &  6.2  &  0.0  &  1.4  &  5.9  &  0.0  &  1.4  &  5.5  &  0.0  &  1.3  &  5.3\\
									\cline{3-16}

				&					& \multirow{2}{*}{0.0}		& 0.0    &  0.0  &  2.6  &  8.3  &  0.0  &  2.3  &  7.7  &  0.0  &  2.3  &  7.2  &  0.0  &  2.1  &  6.7\\
				&					& 					& -0.22  &  0.0  &  2.6  &  8.0  &  0.0  &  2.4  &  7.5  &  0.0  &  2.3  &  7.0  &  0.0  &  2.2  &  6.6
\enddata
\tablecomments{Table notes and notation same as in Table \ref{t:ratessolar}.}
\end{deluxetable*}

\begin{turnpage}
\begin{deluxetable*}{cccccccccccccc} 
\centering
\tabletypesize{\scriptsize}
\tablecaption{{{SN IIb relative rates at solar and low metallicities using $10^6$ single and binary stars generated from distributions of \citet{2017ApJS..230...15M}}} \label{t:ratesMD17}}
\tablewidth{0pt}
\tablehead{
\colhead{} & \colhead{} & \multicolumn{6}{c}{$M_{\rm He~core} \geq 0M_\sun$}  & \multicolumn{6}{c}{$2M_\sun \leq M_{\rm He~core} \leq 6M_\sun$} 
\\
\cmidrule(lr){3-8} \cmidrule(lr){9-14} 
\colhead{} & \colhead{} & \multicolumn{3}{c}{$M_{\rm H~env} \leq 1.0M_\sun$} & \multicolumn{3}{c}{$M_{\rm H~env} \leq 0.5M_\sun$} & \multicolumn{3}{c}{$M_{\rm H~env} \leq 1.0M_\sun$} & \multicolumn{3}{c}{$M_{\rm H~env} \leq 0.5M_\sun$} 
\\
\cmidrule(lr){3-5} \cmidrule(lr){6-8} \cmidrule(lr){9-11} \cmidrule(lr){12-14} 
\colhead{$Z$} & \colhead{$\alpha$} & \colhead{single} & \multicolumn{2}{c}{binary} & \colhead{single} & \multicolumn{2}{c}{binary} & \colhead{single} & \multicolumn{2}{c}{binary} & \colhead{single} & \multicolumn{2}{c}{binary}
\\ 
\cmidrule(lr){4-5} \cmidrule(lr){7-8} \cmidrule(lr){10-11} \cmidrule(lr){13-14} 
\colhead{} & \colhead{} & \colhead{} & \colhead{$\epsilon = 0.5$} & \colhead{$\epsilon = 0.1$} & \colhead{} & \colhead{$\epsilon = 0.5$} & \colhead{$\epsilon = 0.1$} & \colhead{} & \colhead{$\epsilon = 0.5$} & \colhead{$\epsilon = 0.1$} & \colhead{} & \colhead{$\epsilon = 0.5$} & \colhead{$\epsilon = 0.1$}
}
\startdata
			& 1.6		& $0.21-0.4$ & $0.12-0.22$ & $0.15-0.29$ & $0.14-0.26$ & $0.09-0.17$ & $0.12-0.22$ & 0.0 & $0.1-0.19$ & $0.12-0.24$ & 0.0 & $0.07-0.14$ & $0.09-0.17$\\ 
$Z_\odot$ 	& 2.3		& $0.18-0.33$ & $0.15-0.28$ & $0.2-0.37$ & $0.11-0.22$ & $0.11-0.21$ & $0.15-0.28$ & 0.0 & $0.13-0.25$ & $0.17-0.32$ & 0.0 & $0.1-0.18$ & $0.13-0.24$\\
			& 3.0		& $0.13-0.24$ & $0.16-0.3$ & $0.22-0.41$ & $0.08-0.15$ & $0.12-0.23$ & $0.17-0.31$ & 0.0 & $0.14-0.27$ & $0.19-0.36$ & 0.0 & $0.11-0.2$ & $0.14-0.26$\\
\cline{1-14}		
			& 1.6		& $0.0$ & $0.81-1.6$ & $2.8-5.3$ & $0.0$ & $0.73-1.4$ & $2.6-5.1$ & 0.0 & $0.67-1.3$ & $2.2-4.3$ & 0.0 & $0.64-1.2$ & $2.2-4.2$\\
$Z_\odot/4$ 	& 2.3		& $0.0$ & $0.8-1.5$ & $2.6-5.0$ & $0.0$ & $0.72-1.4$ & $2.5-4.8$ & 0.0 & $0.67-1.3$ & $2.2-4.2$ & 0.0 & $0.64-1.2$ & $2.2-4.1$\\
			& 3.0		& $0.0$ & $0.83-1.6$ & $2.7-5.0$ & $0.0$ & $0.77-1.4$ & $2.6-4.8$ & 0.0 & $0.74-1.4$ & $2.3-4.4$ & 0.0 & $0.71-1.3$ & $2.3-4.2$	
\enddata
\tablecomments{Table notes and notation same as in Table \ref{t:ratessolar}. The lower (upper) limit on rates is from assuming that all (none) of the secondaries explode as SNe.}
\end{deluxetable*}
\end{turnpage}





\bibliography{/Users/test/Dropbox/Work/references-master}
\bibliographystyle{apj}

\end{document}